\documentclass[twocolumn,showpacs,preprintnumbers,amsmath,amssymb]{revtex4}

\usepackage{graphicx}
\usepackage{dcolumn}
\usepackage{bm}
\usepackage{epsf}

\newcommand{\bea}{\begin{eqnarray}}
\newcommand{\eea}{\end{eqnarray}}
\newcommand{\beqa}{\begin{eqnarray}}
\newcommand{\eeqa}{\end{eqnarray}}
\newcommand{\eq}[1]{eq.~(\ref{#1})}

\newcommand{\Eq}[1]{Eq.~(\ref{#1})}

\newcommand{\ur}[1]{(\ref{#1})}
\newcommand{\urs}[2]{(\ref{#1},\ref{#2})}

\newcommand{\beq}{\begin{equation}}
\newcommand{\eeq}{\end{equation}}
\newcommand{\la}[1]{\label{#1}}
\newcommand{\ba}{\begin{array}}
\newcommand{\ea}{\end{array}}

\newcommand{\half}{{\textstyle{\frac{1}{2}}}}

\newcommand{\n}{\nonumber}
\newcommand{\inte}{interference\;}

\begin{document}

\title{To see the exotic $\Theta^+$ baryon from interference}
\author{Moskov Amarian$^{a}$}
\author{Dmitri Diakonov$^{b,c}$}
\author{Maxim V. Polyakov$^{b,c}$}
\affiliation{
$^a$ Old Dominion University, Norfolk, Virginia 22901, USA\\
$^b$ Petersburg Nuclear Physics Institute,
Gatchina, 188 300, St. Petersburg, Russia\\
$^c$ Institut f\"ur Theoretische Physik II, Ruhr-Universit\"at Bochum, Bochum
D-44780, Germany}

\date{December 12, 2006}

\begin{abstract}
Since all couplings of the exotic $\Theta^+$ baryon to normal hadrons seem to
be small it is hard to reveal it in standard resonance searching. We suggest
to look for the $\Theta^+$ production in interference with a known resonance
yielding the same final state but having a high production rate. The
interference process is linear in the $\Theta^+$ couplings whereas the
non-interference process is quadratic. That gives an obvious gain if the couplings are
small. Moreover, employing the peculiar oscillating nature of the interference
processes one can reduce considerably the parasitic background and determine the
$\Theta^+$ resonance parameters.
\end{abstract}


\maketitle

\section{Introduction}

The original observation of a narrow exotic baryon resonance in two
independent experiments by T.~Nakano {\it et al.}~\cite{Nakano-1} and
A.~Dolgolenko {\it et al.}~\cite{Dolgolenko-1}, announced in the end of 2002~\cite{footnote-1}
were followed in 2003-04 by a dozen experiments confirming the resonance and about
the same amount of non-sighting experiments. In 2005 the results of the two CLAS
high-statistics $\gamma d$ and $\gamma p$ experiments were
announced~\cite{CLAS-d2,CLAS-d2Lambda,CLAS-p2}, which didn't see any statistically
significant signal of the $\Theta^+$ resonance and gave upper bounds for its production
cross sections. Although those upper bounds didn't contradict the theoretical
estimates~\cite{K*exchange,Titov,Guzey}, see also~\cite{D-06}, many people
in the community jumped to the conclusion that ``pentaquarks do not exist".
This conclusion is premature as in 2005-06 new results became
available~\cite{Nakano-2,Dolgolenko-2,SVD-2} partly based on new data, confirming
seeing the $\Theta^+$.

In particular, A.~Dolgolenko {\it et al.}~\cite{Dolgolenko-2} have doubled the
statistics of their $K^+n({\rm Xe})\to K^0p$ data as compared to the original
DIANA experiment~\cite{Dolgolenko-1}. Previously, there were about 30 events
attributed to the $\Theta^+$ resonance, now there are about 60, as it should be if
the signal is real. Also, more thorough analysis has been performed to
understand better the kinematics of the reaction and the background processes.
The resonance peak is seen already in the raw data but is strongly enhanced by
a mild kinematical cut suppressing re-scattering. The authors estimate the
statistical significance of the resonance as $S/\sqrt{B}=7.3\,\sigma$. The mass
is found to be $m_{\Theta}=1537\pm 2\,{\rm MeV}$ and the width
$\Gamma_{\Theta}=0.36\pm 0.11\,{\rm MeV}$ (with possible systematic
uncertainties). This is the only experiment where the direct estimate of the
width is possible since the {\em formation} cross section integrated over the
resonance range is proportional to the width~\cite{CT}. The only other
available formation experiment with the secondary kaon beam at BELLE sets an
upper limit $\Gamma_\Theta<0.64\,{\rm MeV}$ (at a 90\% confidence
level)~\cite{BELLE} which is beyond the above value. We remark
that the reanalysis of the old $KN$ scattering data~\cite{Arndt:2003xz}
shows that there is room for the exotic resonance with a width below 1~MeV.

The small width implies that the coupling $g_{\Theta NK}$ is at least an order
of magnitude less than $g_{NN\pi}\!\approx\!13$. The small value of $g_{\Theta NK}$
appears naturally in a relativistic field-theoretic approach to baryons,
allowing for a consistent account for multi-quark components in
baryons; in particular in Ref.~\cite{DP-05,Lorce} an upper bound
$\Gamma_\Theta\approx 2\,{\rm MeV}$ has been obtained without any parameter
fixing. An even smaller width comes out from the parameter-free QCD sum rules
analysis~\cite{Oganesian}. As to the $\Theta^+$ coupling to the vector $K^*$ meson,
its `electric' part corresponding to the $\gamma_\mu$ vertex is anyway very small as it
vanishes at zero momentum transfer in the $SU(3)$-symmetric limit, and its
`magnetic' part corresponding to $\sigma_{\mu\nu}q_\nu$ is proportional to the
$\Theta N$ transition magnetic moment which is expected to be an order of
magnitude less than the nucleon magnetic moments~\cite{Polyakov-magnetic}. In
fact, all $\Theta^+$-nucleon-meson couplings vanish in the imaginary non-relativistic
limit when ordinary baryons are made of three quarks only with no admixture
of $Q\bar Q$ pairs. This was the base for the prediction of a narrow pentaquark
in the first place~\cite{DPP-97}.

If indeed all $\Theta^+$-nucleon-meson transition amplitudes are
as small as are expected, it becomes a non-trivial task to reveal
$\Theta^+$ in {\em production} (as contrasted to formation) experiments.
For example, in the recent study of the $\gamma p\to K^0\bar K^0p$ reaction by
CLAS collaboration no statistically significant resonance structure was
observed despite record statistics, and only an upper limit of the
$\Theta^+$ production cross sections of $\sim\!0.7\,{\rm nb}$ was
obtained~\cite{CLAS-p2,footnote-2}. However, a theoretical
estimate performed prior to the experiment and based on the
(Reggeized) $K^*$ exchange with the small transition magnetic
moment mentioned above, gave only $\sim\!0.2\,{\rm nb}$ for that
cross section~\cite{K*exchange}. [Recent phenomenological analysis of the
$K^*$ coupling to the antidecuplet~\cite{Azimov:2006he} has confirmed its
smallness.] It demonstrates how hard it is to make a definite conclusion
about the $\Theta^+$ existence from a production experiment employing a
standard resonance-searching procedure.

In this paper, we suggest to search for the narrow $\Theta^+$ resonance in
a non-standard way, exploiting the interference of the small $\Theta^+$ production
amplitude with the large production amplitude of a known resonance,
yielding the same final state. Although the interference idea is
very general, we apply it primarily to the CLAS experiment~\cite{CLAS-p2}
whose impressive amount of data can be used to look for the $\Theta^+$ resonance
in interference with the large $\phi$ photoproduction, see Fig.~1.
Since the final state in both cases is the same, the two amplitudes {\em must}
interfere unless forbidden kinematically. A simple account for kinematics shows
that the two amplitudes interfere within the photon lab energy range
$1.74<E_\gamma<2.15\,{\rm GeV}$ which is inside the range studied by CLAS.

The $\Theta^+$ production amplitude {\em squared} has been estimated in
Ref.~\cite{K*exchange} with the tiny result for the cross section in the sub-nanobarn
range -- too small to be observable even with the large CLAS statistics.
However, the interference cross section is {\em linear} in the $\Theta^+$ coupling
and hence can be substantially larger. In addition, the \inte cross section
has a peculiar oscillating nature which may help to establish the resonance
mass and width without resolving the Breit--Wigner peak, which is impossible
because it is so narrow.

\section{The $\gamma p\to K^0\bar K^0p$ reaction}

\subsection{$K^0\bar K^0$ {\it versus} $K_{\!L}K_{\!S}$}

In the recent CLAS experiment~\cite{CLAS-p2} studying the above reaction,
$K_{\!S}$ (decaying into $\pi^+\pi^-$) and $p$ have been detected. The second
kaon was reconstructed from the missing mass of all detected particles.
A large portion of events were due to the production of the $\phi$ meson
decaying into $K_{\!L}K_{\!S}$. These events have been rejected in the previous
analysis~\cite{CLAS-p2}, however they are exactly what are needed now: Fig.~1, right.
Since $\phi$ is a vector meson, the amplitude is antisymmetric under the
interchange of $K_{\!L}$ and $K_{\!S}$.

\begin{figure}[htb]
\includegraphics[width=0.45\textwidth]{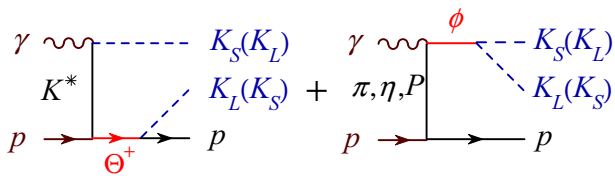}
\caption{Two $\gamma p$ processes producing the $K_{\!L}K_{\!S}p$ final state: via
the $\Theta^+$ resonance (left) and via the $\phi$ resonance (right).}
\end{figure}

Since $\Theta^+$ has strangeness +1 it necessarily decays into $K^0$, hence
the upper line in Fig.~1, left, is $\bar K^0$. Using $K^0=(K_{\!S}+K_{\!L})/\sqrt{2}$,
$\bar K^0=(K_{\!S}-K_{\!L})/\sqrt{2}$ (we neglect the one-per-mill effect of CP violation)
one rewrites the $(\bar K^0K^0)$ production amplitude via $\Theta^+$ as
$\half(K_{\!S}K_{\!S})-\half(K_{\!L}K_{\!L})+\half(K_{\!S}K_{\!L})-\half(K_{\!L}K_{\!S})$.
Only the last two terms interfere with the above amplitude of $\phi$ production,
and we see that they are also antisymmetric under the interchange of $K_{\!L}$ and $K_{\!S}$.
The two antisymmetric amplitudes can and do interfere; the resulting
interference cross section is symmetric under $K_{\!L}\!\leftrightarrow\!K_{\!S}$.

\subsection{Kinematics}

The $2\to 3$ reaction is characterized, at given values of masses of final
particles, by 5 invariants, one of them being the invariant reaction energy
$\surd{s}$. In this case $s=m_N^2+2m_NE_\gamma$ where $E_\gamma$ is the energy
of the incoming photon in the proton rest (or lab) frame. The other two
invariants can be chosen to be the invariant masses of $K_{\!L}K_{\!S}$
(call it $m_{12}$) and of $K_{\!S}p$ (call it $m_{23}$). The last two
invariants can be chosen more or less arbitrarily. The phase volume
of the reaction in the $(m_{12},m_{23})$ axes is shown, for various
photon energies, in Fig.~2.

\begin{figure}[htb]
\includegraphics[width=0.22\textwidth]{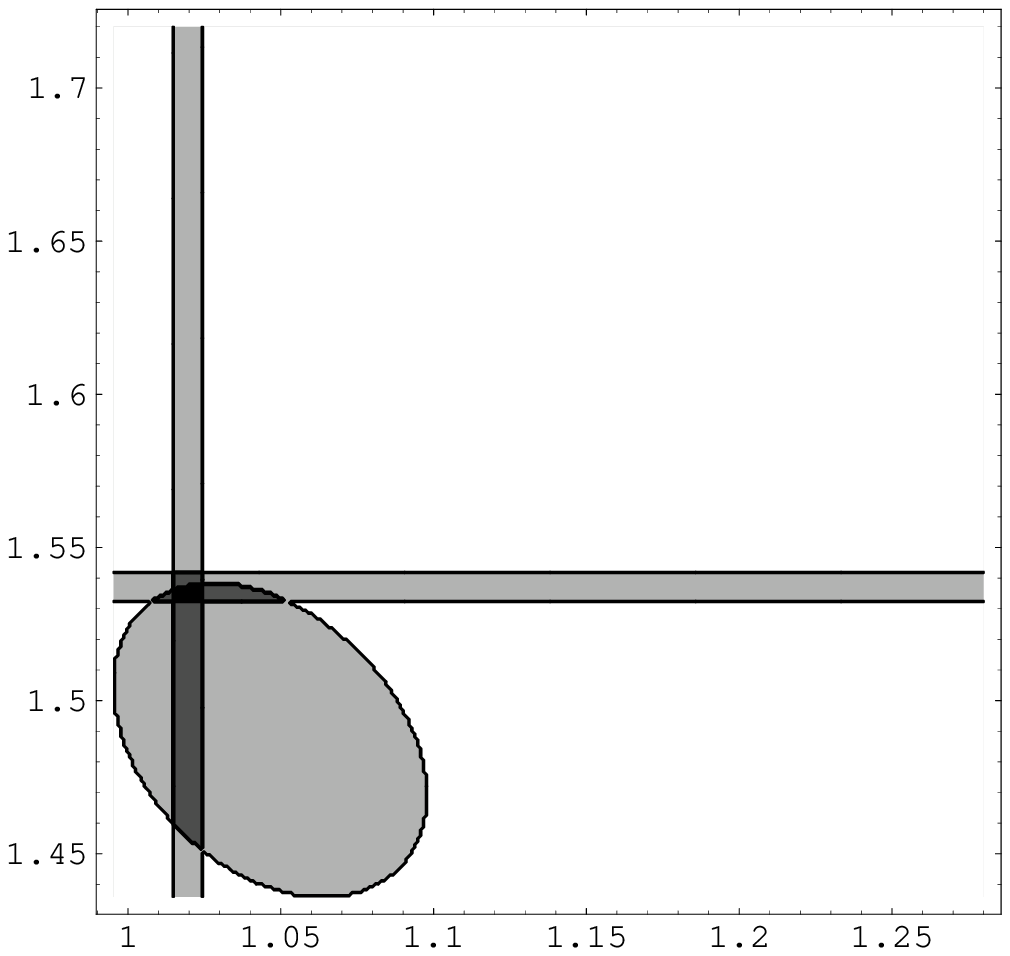} 
\includegraphics[width=0.22\textwidth]{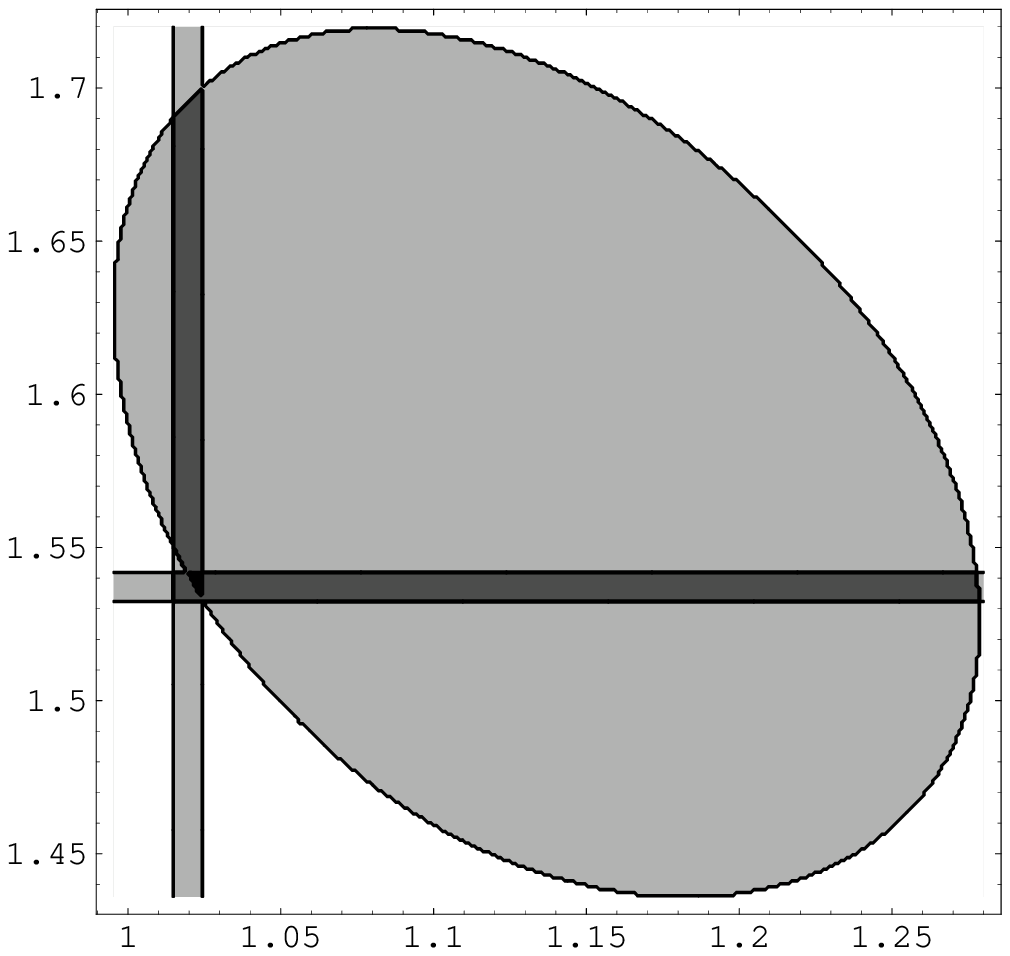} 
\includegraphics[width=0.22\textwidth]{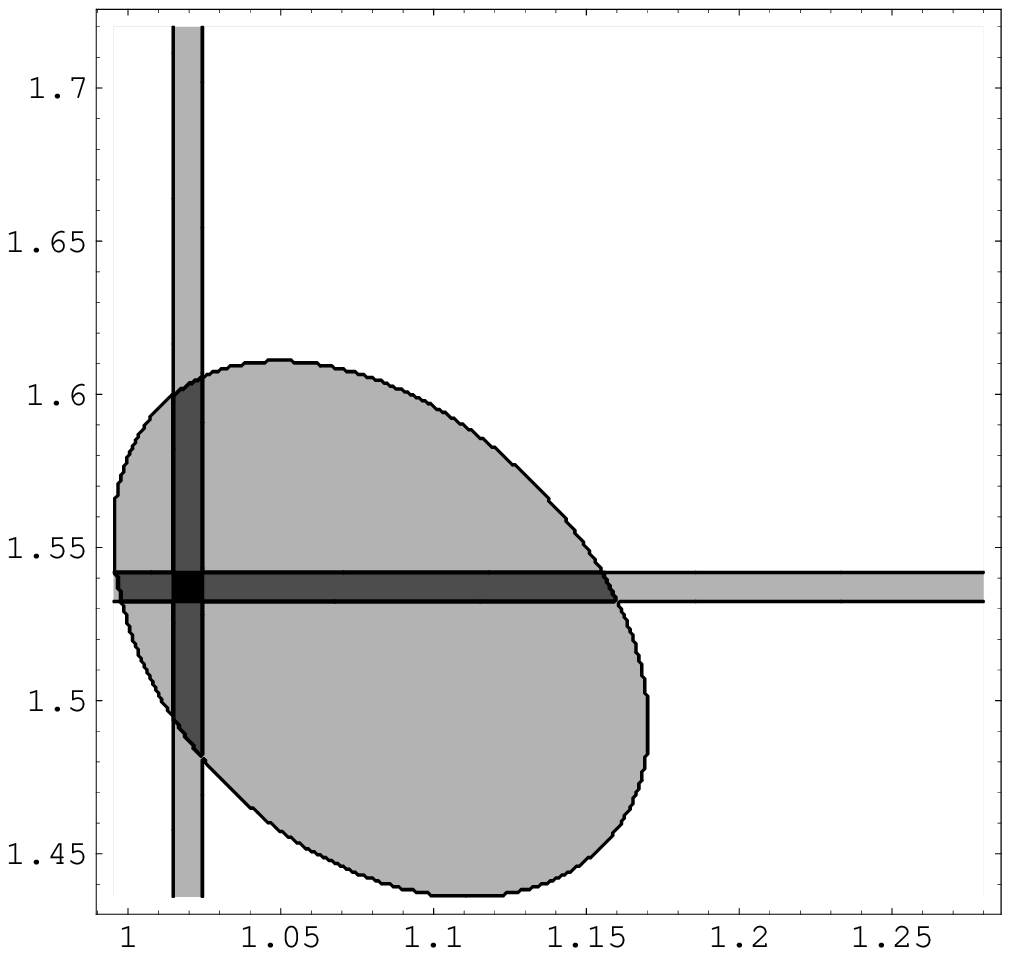} 
\caption{Phase volumes (Dalitz plots) for the $\gamma p\to K_{\!L}K_{\!S}p$
reaction at photon lab energies $E_\gamma=1.74,\,2.15$ and $1.90\,{\rm GeV}$.
The $K_{\!L}K_{\!S}$ invariant mass $m_{12}$ is plotted along the horizontal axis,
and the $K_{\!S}p$ invariant mass $m_{23}$ is plotted along the vertical axis
(note the linear scale). The strips show the positions of the $\phi$ and
$\Theta^+$ resonances. The $\Theta\!-\!\phi$ interference occurs at the
intersection of the strips, thus within the range of $E_\gamma$ from $1.74$
to $2.15\,{\rm GeV}$ (bottom).}
\end{figure}

Energies below 1.74 GeV are too small to produce the $\Theta^+$ resonance
(whose mass we assume equal 1537 MeV); at energies above 2.25 GeV the decay
products of $\Theta^+$ and $\phi$ do not overlap, hence there is no
interference.

\subsection{A sum of two Breit--Wigner's}

In the $\gamma p\to K_{\!L}K_{\!S}p$ amplitude, there are two rapidly varying
functions: one is the $\phi$ resonance pole in the invariant variable $m_{12}$
and the other is the $\Theta^+$ resonance pole in the invariant variable
$m_{23}$. The corresponding widths, $\Gamma_\phi=4.26\,{\rm MeV}$ and
$\Gamma_\Theta\sim 0.5\,{\rm MeV}$, are very small as compared to the typical
hadron masses of several hundred MeV which define the scale of variation of other
factors in the amplitude. Therefore, we can write the fast varying pieces
of the amplitude without knowing the detailed dynamics of the process. It is
a coherent sum of two Breit--Wigner amplitudes in the variables $m_{12}$ and
$m_{23}$, times slowly varying factors:
\beq
{\cal A}=A_\phi\frac{\sqrt{\Gamma_\phi}}{m_{12}-m_\phi+i\frac{\Gamma_\phi}{2}}
+A_\Theta\frac{\sqrt{\Gamma_\Theta}}{m_{23}-m_\Theta+i\frac{\Gamma_\Theta}{2}}
+B
\la{ampl1}\eeq
where $A_{\phi,\Theta}$ are the $\phi,\Theta$ resonance production amplitudes
which are, generally, functions of all kinematical invariants but are slowly
varying on the scale of $\Gamma_{\phi,\Theta}$. Having a dynamical model for
the resonance production these amplitudes can be computed. We shall, however,
attempt to extract as much information from \eq{ampl1} as possible without
knowing their detailed form. We have added a non-resonance amplitude $B$
which does not contain the $s$-channel resonances and hence is also a slowly
varying function of $m_{12}$ and $m_{23}$. $A_{\phi,\Theta}$ and $B$ are
generally complex. We introduce the ratios
\beq
\frac{A_\Theta}{A_\phi}:=R\,e^{i\delta},\qquad
\frac{B}{A_\phi}:=\rho\,e^{i\eta}\,.
\la{ratios}\eeq
To shorten equations, we introduce also the shifts of invariant masses
from their resonance positions:
\beq
\Delta_\phi:=2(m_{12}-m_\phi),\qquad \Delta_\Theta:=2(m_{23}-m_\Theta).
\la{Delta}\eeq

The cross section is proportional to $|{\cal A}|^2$ which we write as
\bea\n
&&\sigma(m_{12},m_{23},...)=|C|^2\left[\frac{\Gamma_\phi}{\Delta_\phi^2+\Gamma_\phi^2}
+R^2\frac{\Gamma_\Theta}{\Delta_\Theta^2+\Gamma_\Theta^2}+\frac{\rho^2}{4}\right.\\
\n\\
\n
&&+2R\sqrt{\Gamma_{\!\!\phi}\Gamma_\Theta}
\frac{(\!\Delta_\phi\Delta_\Theta\!+\!\Gamma_{\!\!\phi}\Gamma_\Theta\!)\cos\delta\!
+\!(\!\Delta_\phi\Gamma_\Theta\!-\!\Gamma_{\!\!\phi}\Delta_\Theta\!)\sin\delta}
{(\Delta_\phi^2+\Gamma_\phi^2)(\Delta_\Theta^2+\Gamma_\Theta^2)}\\
\n\\
\n
&&+\rho\sqrt{\Gamma_\phi}\frac{\Delta_\phi\cos\eta-\Gamma_\phi\sin\eta}
{\Delta_\phi^2+\Gamma_\phi^2}\\
\n\\
&&\left.+R\rho\sqrt{\Gamma_\Theta}\frac{\Delta_\Theta\cos(\eta\!-\!\delta)
-\Gamma_\Theta\sin(\eta\!-\!\delta)}
{\Delta_\Theta^2+\Gamma_\Theta^2}\right]\,.
\la{sigma1}\eea
The first term is the square of the $\phi$ resonance production amplitude,
the second term is the square of the $\Theta$ resonance production amplitude
(suppressed by the {\em square} of the small ratio of the two amplitudes $R^2\ll
1$), the third term is the non-resonant contribution. In principle, one has to
add to it a possible non-coherent non-resonant contribution but such terms will
be irrelevant for our procedure, below.

The most interesting term is the second line in \eq{sigma1}: it gives
the \inte between the $\phi$ and $\Theta$ resonance amplitudes. The third and
fourth lines are the interference terms of $\phi$ and $\Theta$ respectively with
the non-resonant amplitudes. These terms are non-negligible along the resonance
strips in Fig.~2. The \inte term in the second line is essential at the
intersection of those strips. Let us discuss it in more detail.

First of all, it is linear in the $\Theta$ production amplitude $R$ (relative
to that of the $\phi$ production); at $R\ll 1$ it may be much larger than the
incoherent $\Theta^+$ production. To get an idea how small $R$ is we estimate
it roughly from the ratio of the $\phi$ and $\Theta$ photoproduction. The first
is about $3\,\mu{\rm b}$ at $E_\gamma\approx 2\,{\rm GeV}$ and the second
is about $0.2\,{\rm nb}$~\cite{K*exchange}, which gives an estimate $R^2\sim
1/1500$, $R\sim 1/40$.

Second, the \inte term is proportional to $\sqrt{\Gamma_\phi\Gamma_\Theta}$.
It reflects the fact that if one (or both) of the interfering resonances is
almost stable ($\Gamma\to 0$), its decay products are carried out far away
from the reaction vertex, and there is no interference.

Most important, if one looks for the events where
the $K_{\!L}K_{\!S}$ mass $m_{12}$ is within the $\phi$ resonance width,
meaning $\Delta_\phi\sim\Gamma_\phi$, and where the $K_{\!S}p$ mass $m_{23}$
is within the $\Theta^+$ width, meaning $\Delta_\Theta\sim\Gamma_\Theta$,
the \inte term is of the order $R/(\sqrt{\Gamma_\phi\Gamma_\Theta})$ which may
be quite large despite the smallness of the production rate $R$.
It is helpful that $\phi$ is a narrow resonance.

The \inte term in the second line of \eq{sigma1} depends essentially on the
relative phase $\delta$ of the $\phi$ and $\Theta^+$ production amplitudes.
Can anything be said about $\delta$ without going into a detailed dynamical
model for the amplitudes? As we argued in the Introduction, the $\Theta^+$
production amplitude (Fig.~1, left) is dominated by the Reggeized $K^*$
exchange~\cite{K*exchange}. At low energies $E_\gamma\sim 2\,{\rm GeV}$ we are
interested in, the effect of the Reggeization is not large and it can be fairly
well replaced by the usual $K^*$ meson exchange. The amplitude $A_\Theta$ is then real.
As to the photoproduction of $\phi$ (Fig.~1, right), it is notoriously complicated
at $E_\gamma\sim 2\,{\rm GeV}$, with no single dominating mechanism~\cite{Titov-05}.
It may be a mixture of Pomeron, $\pi^0,\eta$ and other meson exchanges. Out of
these, only the diffractive Pomeron amplitude is nearly purely imaginary as
it is the ``shadow" from the total $\gamma p$ cross section with many particles
produced, at least at high energies where the Pomeron exchange dominates. At 2 GeV,
however, the $\phi$ photoproduction is still far from its asymptotics at high
energies~\cite{Titov-05} and hence the Pomeron exchange cannot dominate.
Therefore, we expect that the amplitude $A_\phi$ is nearly real, too. We, thus,
expect the relative phase $\delta\approx 0$ in the energy range of interest.
Non-zero values of $\delta$ are, however, not excluded until measured directly.

Putting for simplicity $\delta=0$ in \eq{sigma1} we see that the
$\phi\!-\!\Theta$ \inte term is proportional to
\beq
\sigma_{\rm interf}\sim
\frac{\Delta_\phi\Delta_\Theta+\Gamma_\phi\Gamma_\Theta}
{(\Delta_\phi^2+\Gamma_\phi^2)(\Delta_\Theta^2+\Gamma_\Theta^2)}\,.
\la{inter1}\eeq
We note that the \inte term falls off as $1/\Delta_\phi\Delta_\Theta$
at large distances from the resonances positions and it is maximal when
$m_{12}\approx m_\phi$ {\em and} $m_{23}\approx m_\Theta$, {\it i.e.} near the
intersection of the two resonance strips on the Dalitz plot of Fig.~2.
The r.h.s. of \eq{inter1} is positive when both $\Delta_\phi$ and $\Delta_\Theta$ have
the same sign, {\it i.e.} when the invariant masses $m_{12}$ and $m_{23}$ are both above
or both below the centers of the resonances. It means seeing more events in the
upper right and lower left corners of the intersection of two resonance strips.
When $\Delta_\phi$ and $\Delta_\Theta$ have opposite signs, the \inte term may become
negative, meaning seeing less events in the upper left and lower right corners.
This ``checker board'' pattern of events is illustrated in Fig.~3a where \eq{inter1}
is plotted in the $(\Delta_\phi,\Delta_\Theta)$ axes measured in units of the corresponding
widths. Although we gave an argument that in this case $\delta\approx 0$ we also plot in Fig.~3
the excess/deficiency of the number of events from the interference term in \eq{sigma1}
for other values of the relative phase $\delta=\frac{\pi}{2},\,-\frac{\pi}{2}$ and $\pi$.

\begin{figure}[htb]
\begin{minipage}[t]{.49\textwidth}
\includegraphics[width=0.48\textwidth]{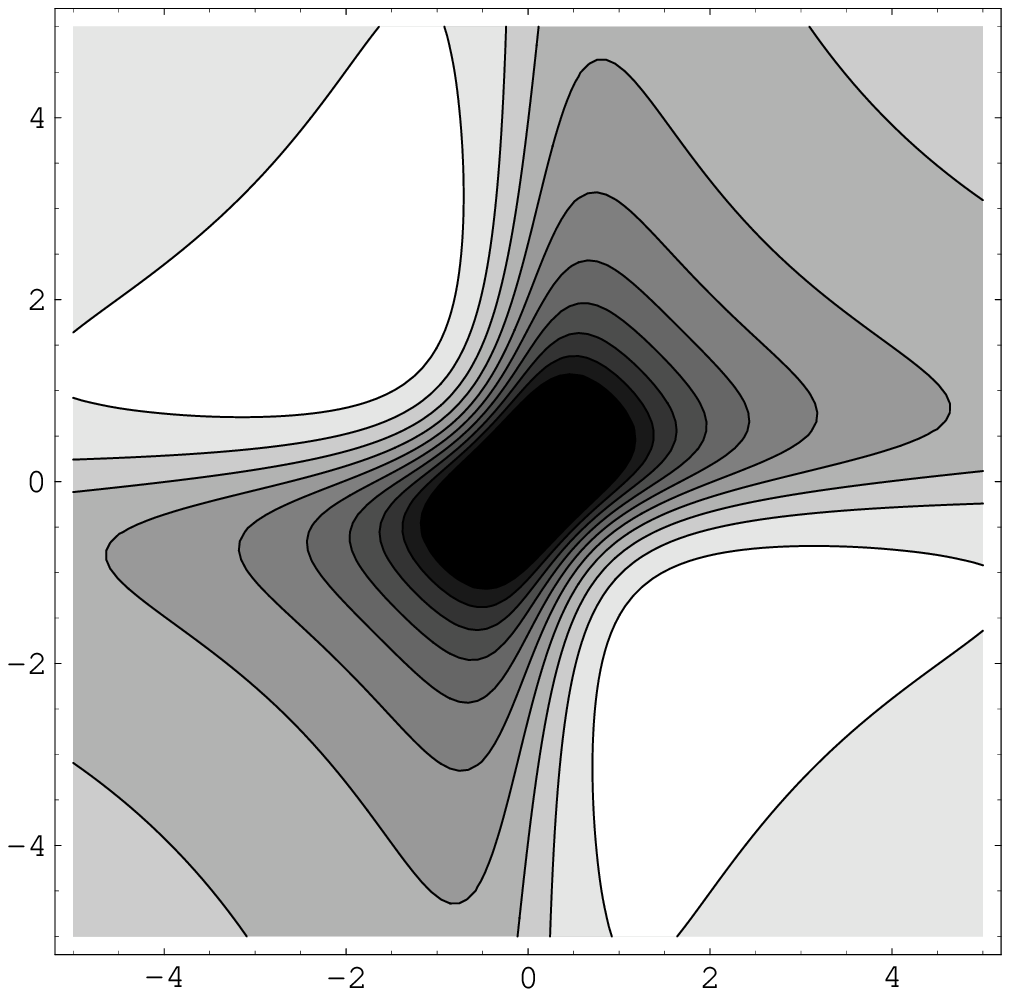} 
\includegraphics[width=0.48\textwidth]{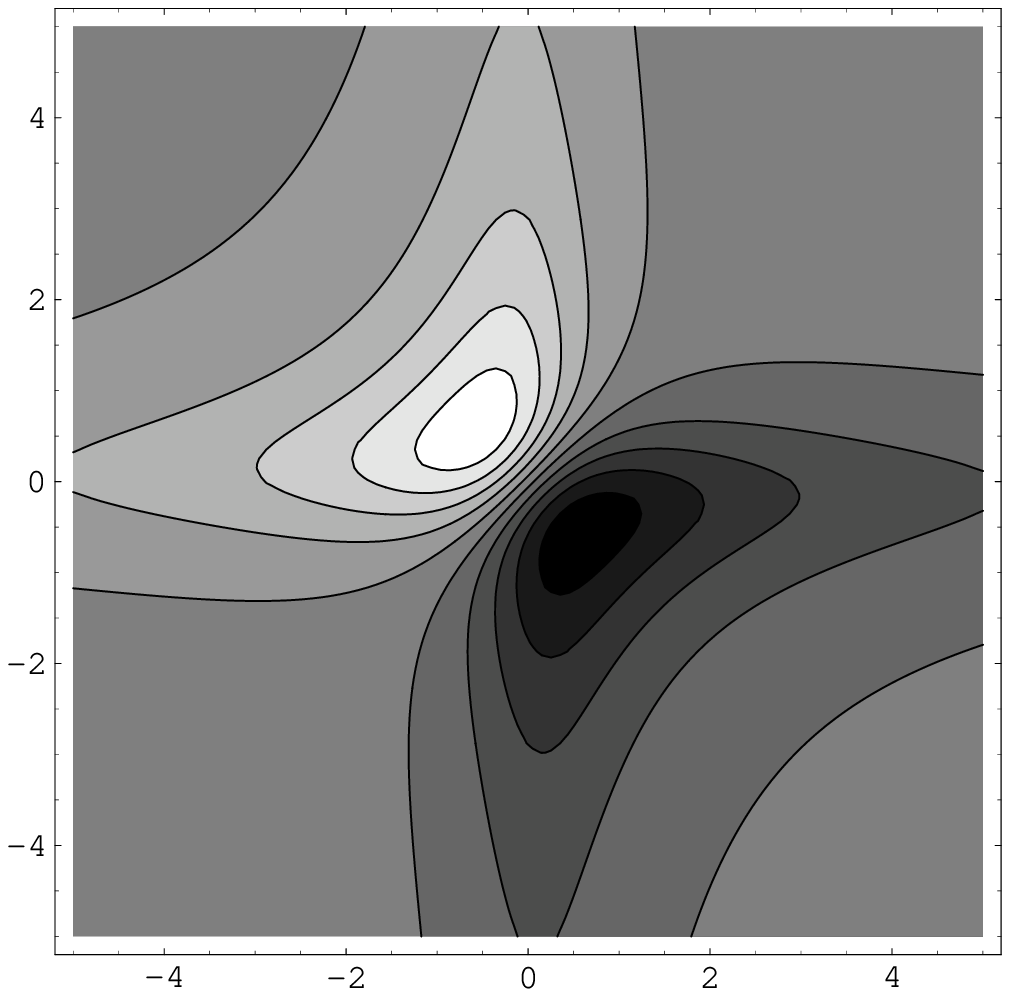} 
\includegraphics[width=0.48\textwidth]{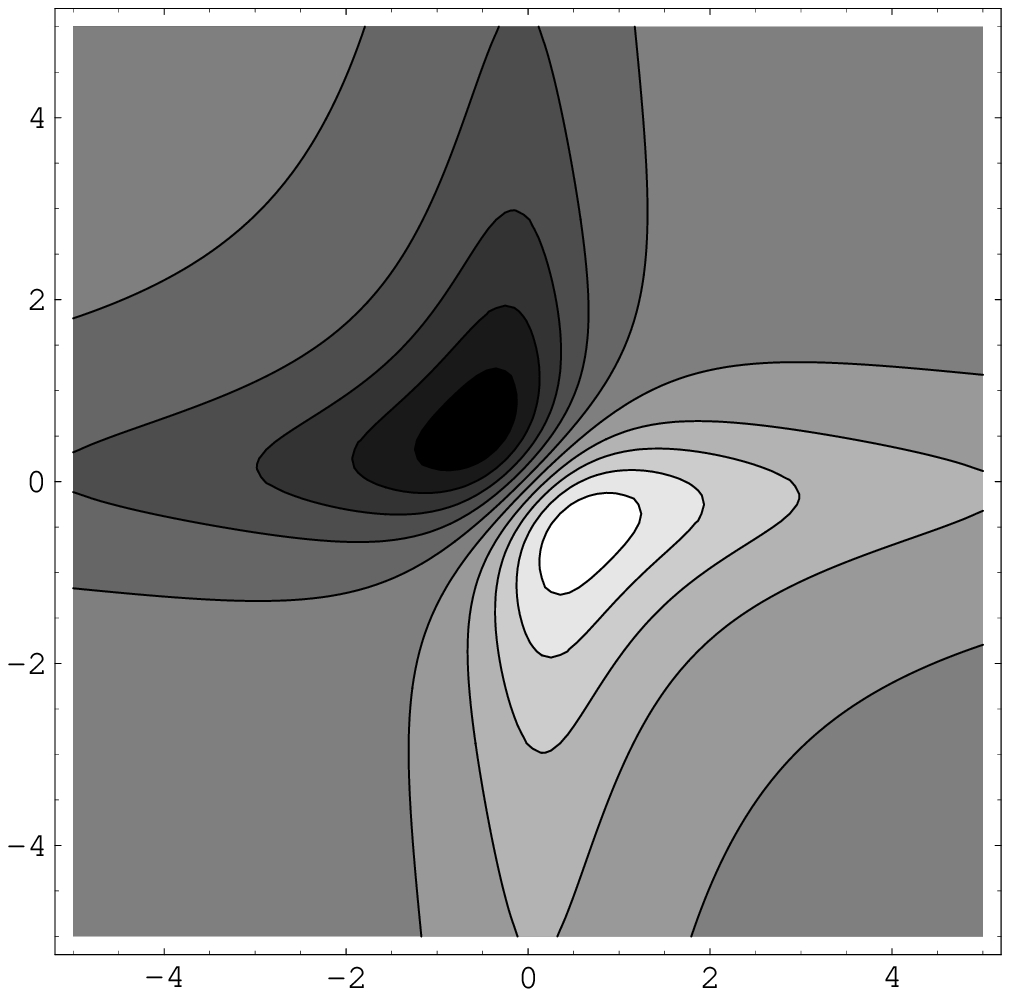}  
\includegraphics[width=0.48\textwidth]{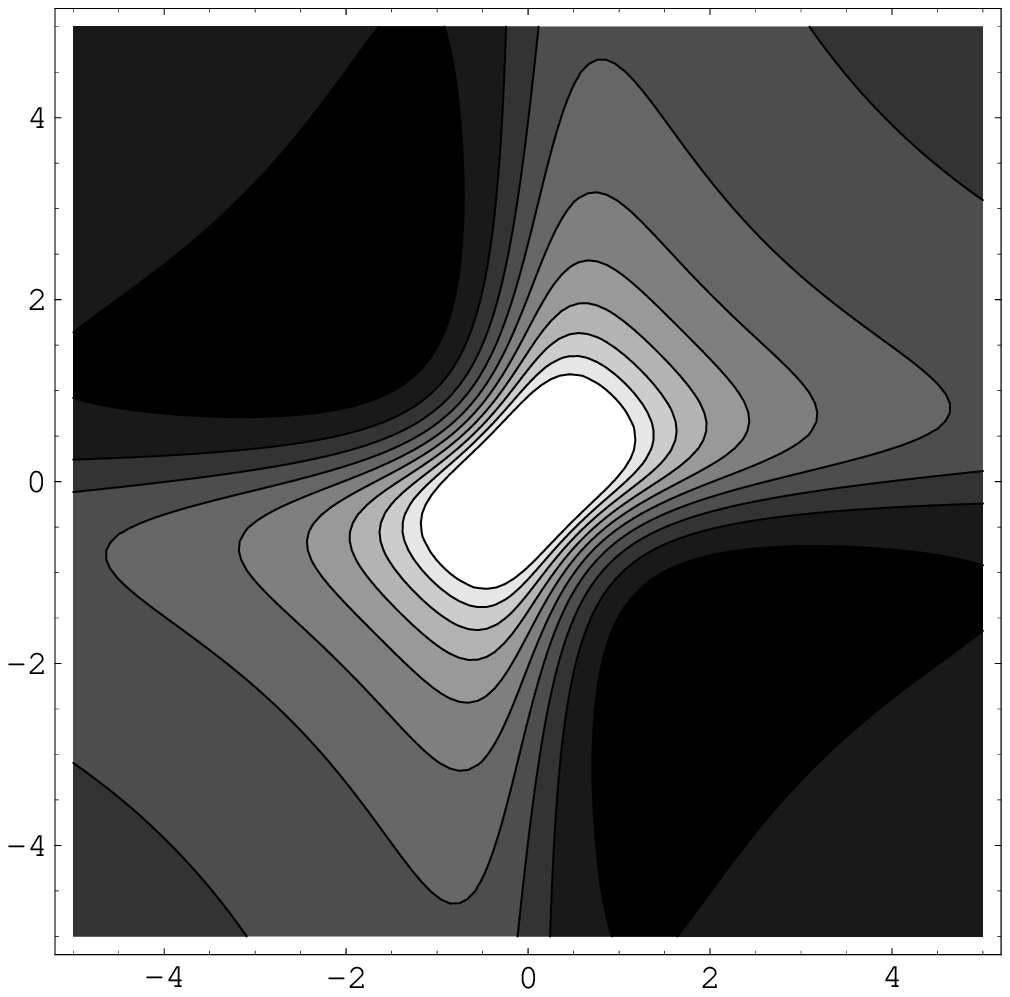}  
\caption{The excess/deficiency of events on the $(m_{12},m_{23})$ Dalitz plot around
the centers of both resonances, at various values of the relative phase
$\delta=0,\frac{\pi}{2},-\frac{\pi}{2},\pi$. These are contour plots with the excess
events shown darker and the deficiency of events shown lighter.}
\end{minipage}
\end{figure}

The difference in the event pattern is so strong that one would be probably
able to say from the first glance on the Dalitz plot what is the relative phase.

\section{Tailoring out the $\Theta^+$ resonance}

The cross section of the $\gamma p\to K_{\!L}K_{\!S}p$ reaction \ur{sigma1} can
be divided into terms even and odd with respect to $\Delta_\phi$, meaning terms
even and odd with respect to the reflection about the $m_{12}=m_\phi$ axis (the
vertical strip in Fig.~2). More interesting information on $\Theta^+$ contains
in the {\em odd} part. To extract it, we integrate all events in some range
of $m_{12}$ from the center of the $\phi$ resonance $m_\phi$ up to some $m_\phi+\mu_\phi$
and {\em subtract} all events integrated symmetrically below the resonance
center, from $m_\phi-\mu_\phi$ to $m_\phi$. The smaller limit $\mu_\phi$
theoretically the cleaner, however one has to make a compromise with the statistics.
Probably, integrating over the resonance within a few units of its width is
optimal. On the one hand, most of the events in the $\phi$
resonance range will be collected and on the other hand, the integration strip
will be still narrow enough to neglect the dependence of other, non-resonance
factors on $m_{12}$.

As the result of this subtraction of events, most terms in \eq{sigma1} cancel
out, most notably the background which is a smooth function at $m_{12}\approx
m_\phi$, and the large square of the $\phi$ production amplitude (the first term
in \eq{sigma1}). We obtain:
\bea\n
&&\int_{m_\phi}^{m_\phi+\mu_\phi}\!\!\!dm_{12}\,\sigma(m_{12},m_{23})
\stackrel{!}{-}\int_{m_\phi-\mu_\phi}^{m_\phi}\!\!\!dm_{12}\,\sigma(m_{12},m_{23})\\
\n\\
\la{sigmaas}
&&=\frac{1}{2}\ln\left[1+\left(\frac{2\mu_\phi}{\Gamma_\phi}\right)^2\right]\\
\n\\
\n
&&\times |C|^2\sqrt{\Gamma_\phi}\left(2R\sqrt{\Gamma_\Theta}\,
\frac{\Delta_\Theta\cos\delta\!+\!\Gamma_\Theta\sin\delta}
{\Delta_\Theta^2+\Gamma_\Theta^2}+\rho\cos\eta\right).
\eea
(we have retained a non-zero $\delta$ for generality).

\Eq{sigmaas} exhibits a $\Theta^+$ resonance term originating from the interference,
and a non-resonant background. Analysis based on \eq{sigmaas} has a clear
advantage over a resonance search based on standard technique. First, the
signal is linear (and not quadratic) in the small production amplitude $R$,
second, most of the parasitic background is canceled by the symmetric
subtraction procedure. Third, the only background left is due to
the {\em coherent} non-resonant production which is not expected to
be large in this case. Fourth, the resonance has a definite oscillating
signature. As we shall see in the next section, it is not likely that
this signature is blurred out by the finite experimental errors.

\section{Smearing with the experimental resolution}

The above equations have been written for an idealized case, assuming the
experimental resolution is perfect. Actually, it never is. Assuming \eq{sigma1}
is written for the `true' values of the variables $m^{\rm true}_{12},\,m^{\rm true}_{23}$
one has to smear it with the probability distribution that the measured
$m^{\rm obs}_{12},\,m^{\rm obs}_{23}$ deviate from the true ones.
We shall assume that the errors in measuring the invariant masses
$m^{\rm obs}_{12}$ and $m^{\rm obs}_{23}$ are statistically independent and
that the error probability distribution is given by the product of two
Gaussian functions with equal widths $\sigma$:
\beq
\frac{\exp\left(-\frac{(m^{\rm obs}_{12}-m^{\rm true}_{12})^2}
{2\sigma^2}\right)}{\sqrt{2\pi\sigma^2}}\cdot
\frac{\exp\left(-\frac{(m^{\rm obs}_{23}-m^{\rm true}_{23})^2}
{2\sigma^2}\right)}{\sqrt{2\pi\sigma^2}}\,.
\la{Gauss1}\eeq
To be concrete, we take the mean square error in measuring $m_{12},\,m_{23}$
to be $\sigma=5\,{\rm MeV}$. If a more realistic error distribution function is
known it should be used instead of \eq{Gauss1}.

To get the observable cross section as function of $m^{\rm obs}_{12}$
and $m^{\rm obs}_{23}$, one has to integrate the theoretical \eq{sigma1} understood as
function of $m^{\rm true}_{12}$ and $m^{\rm true}_{23}$, over these variables with the
error weight \ur{Gauss1}. In \eq{sigma1} there are two kind of functions
encountered: one is symmetric with respect to the resonance center, the other
is  antisymmetric. We introduce their smeared counterparts:
\bea\la{G}
\frac{\Gamma}{\Delta^2+\Gamma^2}&\to&\int\!dm^{\rm true}\,
\frac{\exp\left(-\frac{(m^{\rm obs}-m^{\rm true})^2}
{2\sigma^2}\right)}{\sqrt{2\pi\sigma^2}}\\
\n
\times\!&&\!\!\!\!\!\!
\frac{\Gamma}{4(m^{\rm true}\!-\!m^{\rm res})^2\!+\!\Gamma^2}
:=G(m^{\rm obs}\!-\!m^{\rm res},\Gamma,\sigma),\\
\n\\
\la{D}
\frac{\Delta}{\Delta^2+\Gamma^2}&\to&\int\!dm^{\rm true}\,
\frac{\exp\left(-\frac{(m^{\rm obs}-m^{\rm true})^2}
{2\sigma^2}\right)}{\sqrt{2\pi\sigma^2}}\\
\n
\times\!&&\!\!\!\!\!\!
\frac{2(m^{\rm true}\!-\!m^{\rm res})}{4(m^{\rm true}\!-\!m^{\rm res})^2\!+\!\Gamma^2}
:=D(m^{\rm obs}\!-\!m^{\rm res},\Gamma,\sigma).
\eea
The smeared functions $G$ and $D$ are even (odd) in $m^{\rm obs}\!-\!m^{\rm res}$,
respectively. If the resonance is much more narrow than the experimental
resolution ($\Gamma\ll\sigma$ -- this is the case of the $\Theta^+$ resonance)
the two functions are analytically computable:
\bea\la{Gas}
&&G(\Delta m,\Gamma,\sigma)\stackrel{\Gamma\!\ll\sigma}{=}
\frac{\pi}{2\sqrt{2\pi\sigma^2}}\exp\left(-\frac{(\Delta m)^2}{2\sigma^2}\right),\\
\n\\
\la{Das}
&&D(\Delta m,\Gamma,\sigma)\stackrel{\Gamma\!\ll\sigma}{=}
\frac{\pi}{2\sqrt{2\pi\sigma^2}}\exp\left(-\frac{(\Delta m)^2}{2\sigma^2}\right)\\
\n
&&\cdot {\rm sign}(\Delta m)\,{\rm Erfi}\left(\frac{\Delta m}{\sqrt{2\sigma^2}}\right)
\eea
where ${\rm Erfi}$ is the error function of imaginary argument,
\beq\n
{\rm Erfi}(z)=-i\,{\rm Erf}(iz)=-i\frac{2}{\sqrt{\pi}}\int_0^{iz}\!dt\,e^{-t^2}\,.
\eeq
The two functions $G(\Delta m)$ and $D(\Delta m)$ are plotted in Fig.~4.

\begin{figure}[htb]
\includegraphics[width=0.22\textwidth]{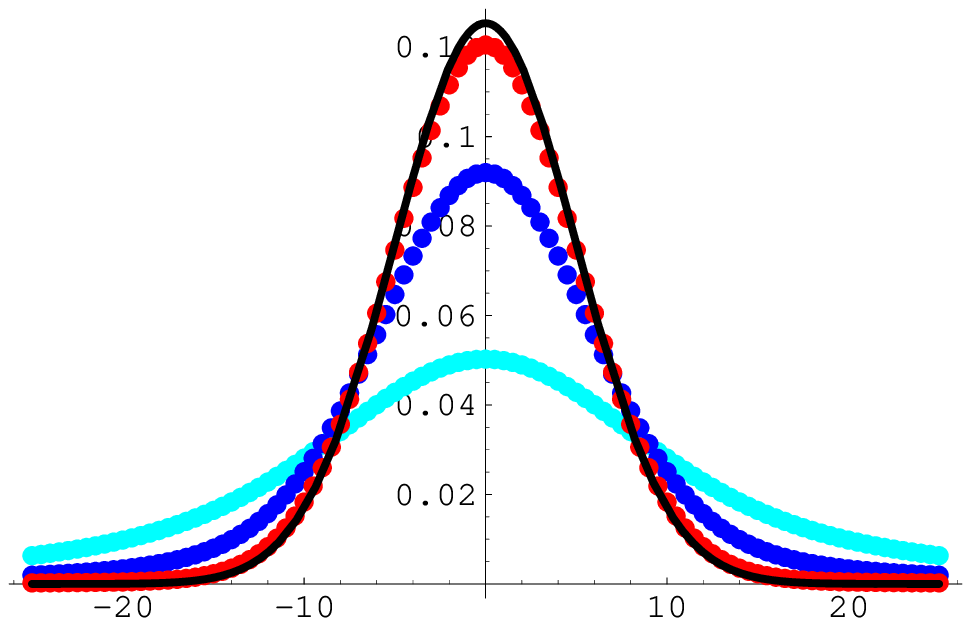} 
\includegraphics[width=0.22\textwidth]{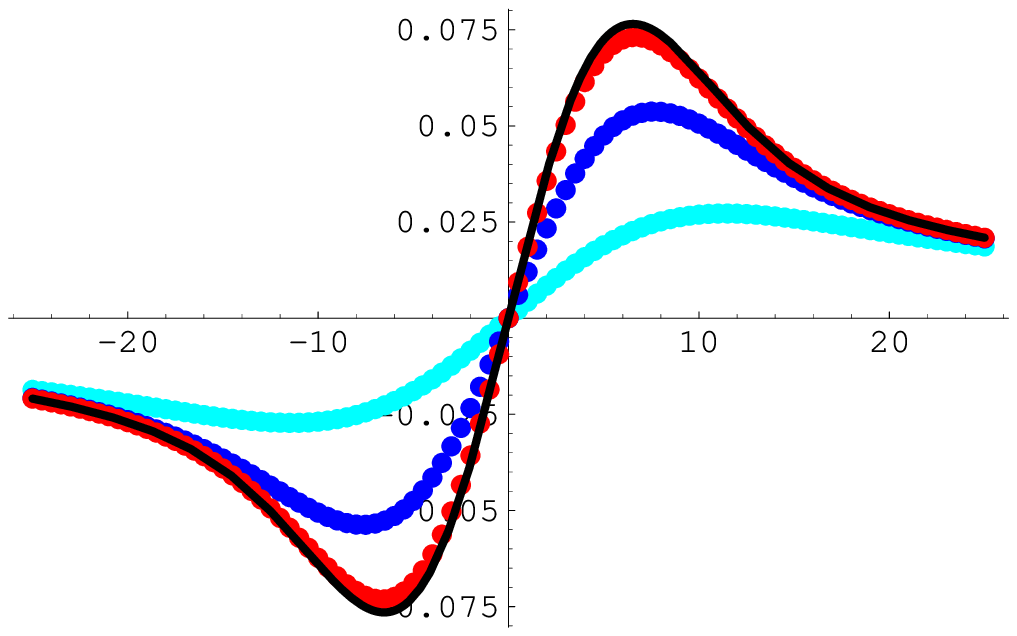} 
\caption{The resonance functions smeared with a Gaussian error distribution, $\sigma=5\,{\rm MeV}$:
the symmetric function $G$ (left) and the antisymmetric function $D$ (right).
The red curve corresponds to smearing the $\Theta^+$ resonance assuming
$\Gamma_\Theta=0.5\,{\rm MeV}$, the blue curve shows the smearing of $\phi$,
$\Gamma_\phi=4.26\,{\rm MeV}$, and the light blue curve shows the smearing of the
$\Lambda(1520)$ resonance, $\Gamma_\Lambda=15.6\,{\rm MeV}$. Solid curves are
the asymptotic $G,D$ functions ($\Gamma\!\to\!0$). The distance to the resonance
$\Delta m$ is plotted along the horizontal axis, in MeV. Note that smearing
of the $\Theta^+$ resonance is well described by the asymptotic functions \urs{Gas}{Das}.}
\end{figure}

It is now easy to write down the cross section resulting from smearing
\eq{sigma1} with the experimental resolution. One has simply to replace the
corresponding factors in \eq{sigma1} by
\bea\n
\frac{\Gamma_\phi}{\Delta_\phi^2+\Gamma_\phi^2}&\to&
G(m_{12}-m_\phi,\Gamma_\phi,\sigma),\\
\n
\frac{\Delta_\phi}{\Delta_\phi^2+\Gamma_\phi^2}&\to&
D(m_{12}-m_\phi,\Gamma_\phi,\sigma),\\
\n
\frac{\Gamma_\Theta}{\Delta_\Theta^2+\Gamma_\Theta^2}&\to&
G(m_{23}-m_\Theta,\Gamma_\Theta,\sigma),\\
\n
\frac{\Delta_\Theta}{\Delta_\Theta^2+\Gamma_\Theta^2}&\to&
D(m_{23}-m_\Theta,\Gamma_\Theta,\sigma).
\eea
All four functions are plotted in Fig.~4. The same replacement should be done
in the $\phi\!-\!\Theta$ interference term, \eq{inter1}, which becomes, after
smearing,
\bea\n
\sigma_{\rm interf}&\sim &
D(m_{12}\!-\!m_\phi,\Gamma_\phi,\sigma)
D(m_{23}\!-\!m_\Theta,\Gamma_\Theta,\sigma)\\
&+&\!\!G(m_{12}\!-\!m_\phi,\Gamma_\phi,\sigma)
G(m_{23}\!-\!m_\Theta,\Gamma_\Theta,\sigma).
\la{inter2}\eea
Its contour plot is shown in Fig.~5. Comparing it with the plot in Fig.~3, left,
we observe that the `checker-board' pattern of the \inte is preserved by
the finite experimental resolution. [It should be reminded however, that
the simple result \ur{inter2} has been obtained assuming that the errors
in measuring the invariant $(K_{\!L}K_{\!S})$ and $(K_{\!S}p)$ masses
are statistically independent. Whether it is indeed a fair approximation
needs a special study.]

\begin{figure}[htb]
\includegraphics[width=0.28\textwidth]{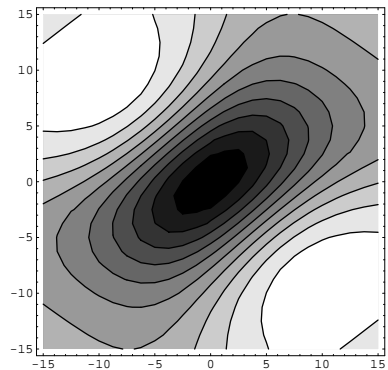} 
\caption{Contour plot for the excess/deficiency of events smeared with the
experimental resolution of 5 MeV, which is due to the $\phi\!-\!\Theta$ interference,
\eq{inter2}. The excess is shown darker and the deficiency shown lighter.
$m_{12}\!-\!m_\phi$ is plotted along the horizontal axis, and $m_{23}\!-\!m_\Theta$
is plotted along the vertical axis, both in MeV. The relative phase $\delta$
is assumed to be zero.}
\end{figure}

We can now virtually apply the $\Theta^+$ identifying procedure described in
Section III. Namely, we integrate all events with $m_{12}$ from the center
of the $\phi$ resonance $m_\phi$ up to some $m_\phi+\mu_\phi$, and {\em subtract}
events integrated from $m_\phi-\mu_\phi$ to $m_\phi$. This procedure nullifies
all terms that are symmetric with respect to the center of the $\phi$ and
stresses terms that are antisymmetric, in particular the \inte term:
\bea\n
&&\int_{m_\phi}^{m_\phi+\mu_\phi}\!\!\!dm_{12}\,\sigma(m_{12},m_{23})
\stackrel{!}{-}\int_{m_\phi-\mu_\phi}^{m_\phi}\!\!\!dm_{12}\,\sigma(m_{12},m_{23})\\
\n\\
\la{sigmaassmea}
&&={\rm const.}\,\left\{2R\sqrt{\Gamma_\Theta}\left[
D(m_{23}-m_\Theta,\Gamma_\Theta,\sigma)\cos\delta\right.\right.\\
\n
&&+\left.\left.
G(m_{23}-m_\Theta,\Gamma_\Theta,\sigma)\sin\delta\right]
+\rho\cos\eta\right\}.
\eea
Apart from a smooth background term $\rho\cos\eta$, \eq{sigmaassmea} exhibits
a characteristic behavior associated with the $\Theta^+$. If the relative phase
$\delta\approx 0$ (as we think it is) only the $D$-function antisymmetric
with respect to the $\Theta^+$ resonance center survives in
\eq{sigmaassmea}; its plot is presented in Fig.~4, right, and is in fact very
close to the asymptotic \eq{Das}. Fitting the difference in the integrated
cross section about the $\phi$ resonance by \eq{sigmaassmea} it is possible
to find the position of the $\Theta^+$ resonance: it is where the
function $D(m_{23}-m_\Theta)$ changes sign.

Determining the value of the width $\Gamma_\Theta$ is more difficult,
especially if the relative phase $\delta\approx 0$. Nevertheless, the shape of
the $D(m_{23}-m_\Theta)$ curve depends implicitly on the width even if
$\Gamma_\Theta$ is less than the experimental resolution $\sigma$, as seen
from the comparison of the curves for $\Gamma=0.5\,{\rm MeV}$ and $4\,{\rm
MeV}$ in Fig.~4, right. Depending on the quality of the data, one would be
probably able to establish from data fitting that $\Gamma_\Theta$ is less
than a few MeV which would be anyway a record achievement.

\section{Triple interference}

The reaction $\gamma p\to K_{\!S}K_{\!L}p$ at $E_\gamma\sim 2\,{\rm GeV}$ is
unique in that all three pairs of the final particles resonate,
and those resonances can interfere when the final states overlap.
Resonance interference can occur when any two of the invariant masses $m_{12}\equiv
m(K_{\!L}K_{\!S})$, $m_{13}\equiv m(K_{\!L}p)$, $m_{23}\equiv m(K_{\!S}p)$, are close
to the resonance masses $m_\phi$, $m_\Theta$, $m_\Theta$, respectively, see
Fig.~6. In fact the distribution in the $m_{13}$ invariant mass has been
also studied in the same experiment~\cite{CLAS-p2} where only an upper
limit for the $\Theta^+$ production has been established.

\begin{figure}[htb]
\includegraphics[width=0.22\textwidth]{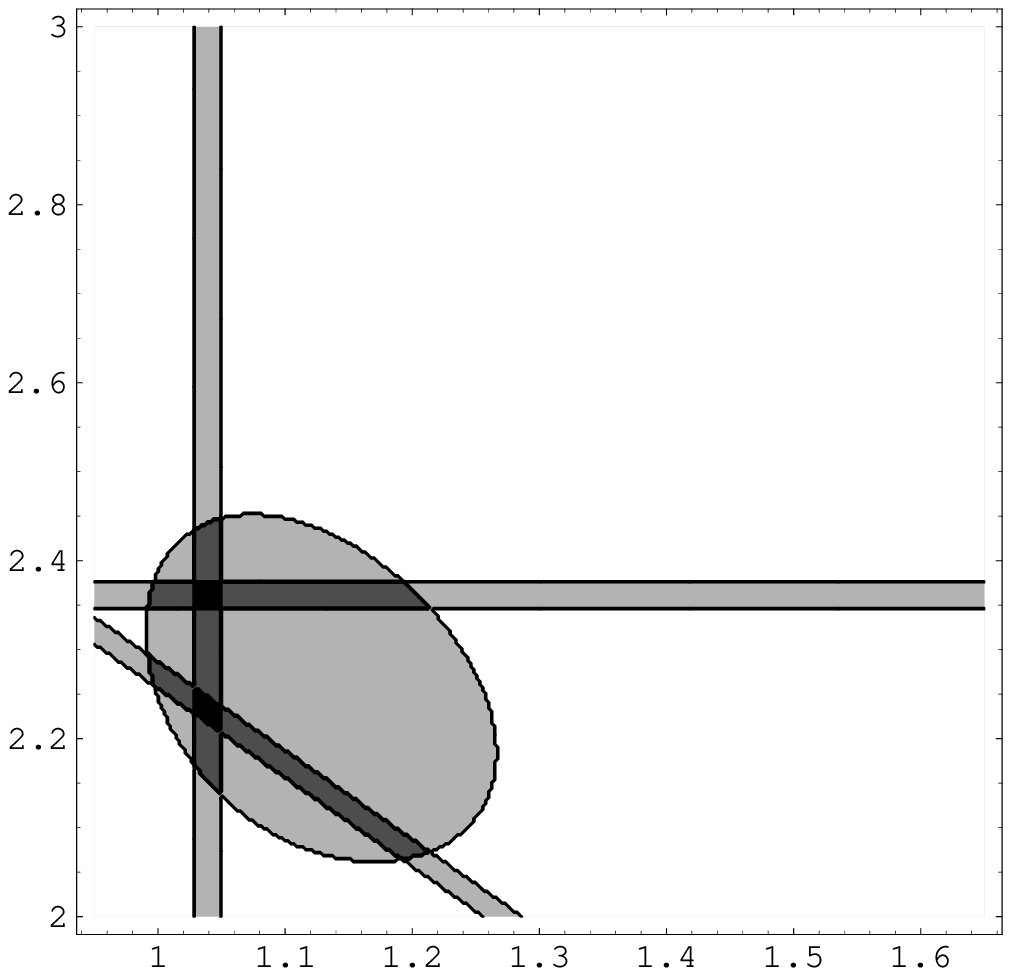} 
\includegraphics[width=0.22\textwidth]{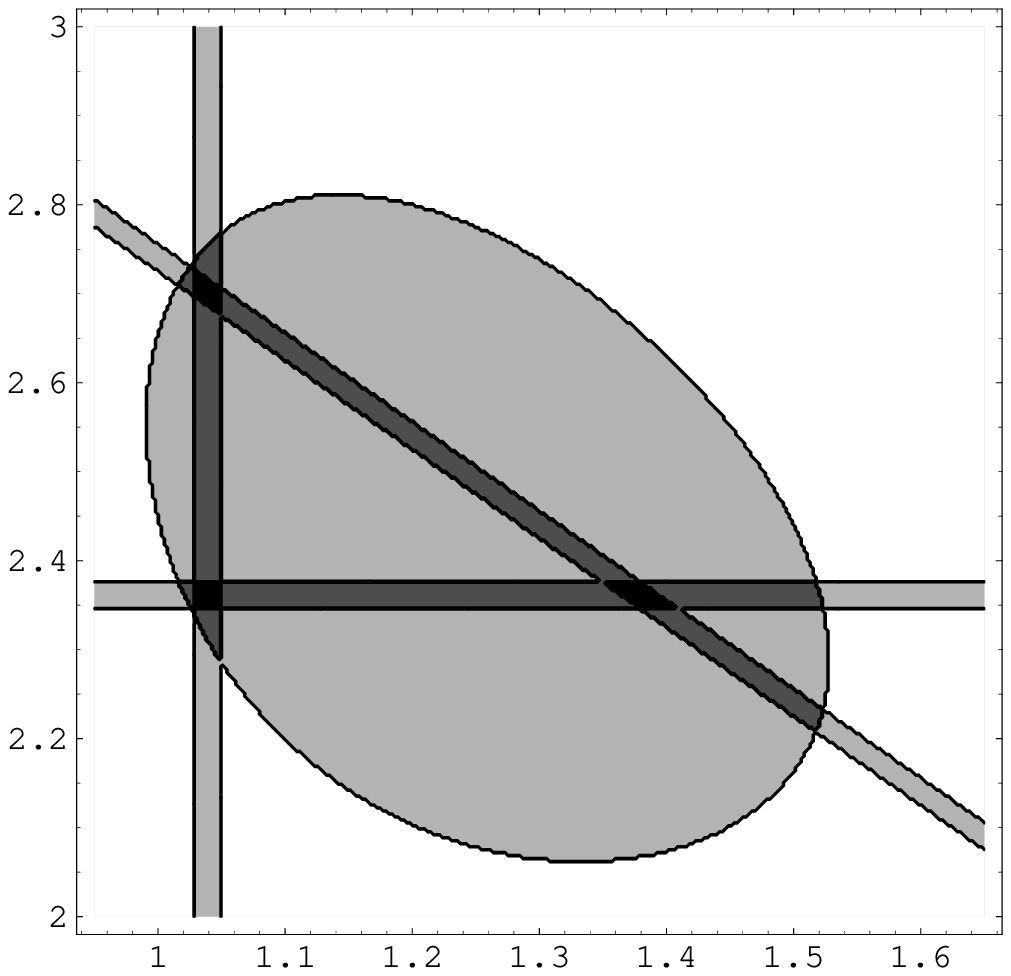} 
\includegraphics[width=0.22\textwidth]{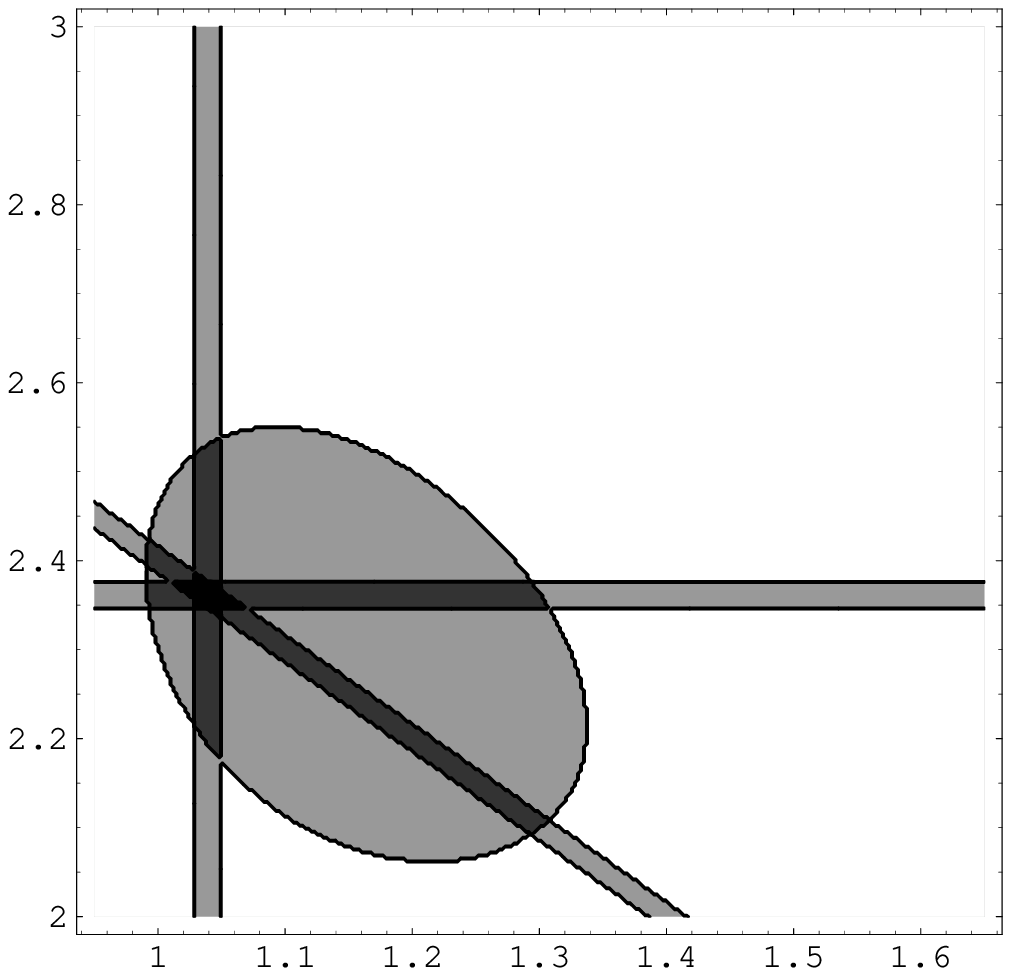} 
\caption{Phase volumes (Dalitz plots) for the $\gamma p\to K_{\!L}K_{\!S}p$
reaction at photon lab energies $E_\gamma=1.80,\,2.05$ and $1.87\,{\rm GeV}$.
The $K_{\!L}K_{\!S}$ invariant mass squared $m_{12}^2$ is plotted along the horizontal axis,
and the $K_{\!S}p$ invariant mass squared $m_{23}^2$ is plotted along the vertical axis.
The strips show the $\phi(K_{\!L}K_{\!S})$, $\Theta(K_{\!S}p)$ and $\Theta(K_{\!L}p)$ resonances.
Resonance interference occurs at the intersection of the strips; at
$E_\gamma=1.87\,{\rm GeV}$ all three resonances interfere (bottom).}
\end{figure}

Everything said above about the $\phi\!-\!\Theta$ interference in the $m_{12}\!-\!m_{23}$
Dalitz plot can be immediately translated into the $\phi\!-\!\Theta$ interference
in the $m_{12}\!-\!m_{13}$ axes. One can apply the event-subtraction method described in
the previous section: it results in the same \eq{sigmaassmea}, with $m_{23}$ replaced
by $m_{13}$. This seems to be a rather cheap way of approximately doubling the statistics
used to analyze \eq{sigmaassmea}.

Another possibility is to look, at $E\gamma\geq 1.85\,{\rm GeV}$, for the interference
of the $\Theta^+$ decay into $K_{\!S}p$ {\it vs} its decay into $K_{\!L}p$.
Introducing the shifts of the invariant masses from the resonance center, $\Delta_\Theta
=2(m_{23}-m_\Theta),\,\Delta_\Theta^\prime=2(m_{13}-m_\Theta)$, one writes the
coherent $\Theta^+$ production cross section at small
$\Delta_\Theta,\Delta_\Theta^\prime$ as
\bea\la{theta-theta}
\sigma^{\Theta\Theta}&\sim &\Gamma_\Theta\left[\frac{|A_\Theta|^2}{\Delta_\Theta^2+\Gamma_\Theta^2}
+\frac{|A^\prime_\Theta|^2}{\Delta_\Theta^{\prime\,2}+\Gamma_\Theta^2}\right.\\
\n\\
\n
&-&\left.2\frac{|A_\Theta||A^\prime_\Theta|(\Delta_\Theta\Delta_\Theta^\prime+\Gamma_\Theta^2)}
{(\Delta_\Theta^2+\Gamma_\Theta^2)(\Delta_\Theta^{\prime\,2}+\Gamma_\Theta^2)}\right]
\eea
where the first two terms stand for the incoherent $\Theta^+$ production observed
through the $K_{\!S}p$ and $K_{\!L}p$ channels, respectively, and the third term
is their interference. $A_\Theta,A^\prime_\Theta$ are the production amplitudes which are,
generally, functions of the invariants but can be replaced by constants when
$m(K_{\!S}p),m(K_{\!L}p)\approx m_\Theta$. In the ideal geometry case
$|A_\Theta|=|A^\prime_\Theta|$, however the experimental acceptance may violate this symmetry.

In this case interference is not amplified by the large $\phi$ production amplitude
and all terms are quadratic in the $\Theta^+$ production amplitude. After
smearing with the experimental resolution (Section IV) the effect of
interference becomes insignificant. Nevertheless, it might be helpful to make
a Dalitz plot of the events in the $m_{13}\!-\!m_{23}$ axes: at the intersection
of the $m(K_{\!S}p)\approx m_\Theta, \,m(K_{\!L}p)\approx m_\Theta$ strips there
can be more events associated with the $\Theta^+$ production than one can discriminate
in the separate $K_{\!S}p$ and $K_{\!L}p$ mass spectra.

Finally, the most non-trivial {\em triple} interference happens at
$E_\gamma\approx 1.87\,{\rm GeV}$ where all three resonance strips cross at one point,
see Fig.~6, bottom. Keeping only the presumably largest terms linear in the
$\Theta^+$ production amplitudes, one generalizes the second line term in \eq{sigma1}
to include interference between the $\phi$ meson and the two $\Theta$'s, one
decaying into $K_{\!S}p$ and the other into $K_{\!L}p$:
\bea\la{sigma3}
&&\sigma^{\phi\Theta\Theta}=2|C|^2\frac{\sqrt{\Gamma_\phi\Gamma_\Theta}}
{\Delta_\phi^2+\Gamma_\phi^2}\\
\n\\
\n
&&\cdot\left[\left(R\frac{\Delta_\phi\Delta_\Theta+\Gamma_\phi\Gamma_\Theta}
{\Delta_\Theta^2+\Gamma_\Theta^2}
+R'\frac{\Delta_\phi\Delta'_\Theta+\Gamma_\phi\Gamma_\Theta}
{\Delta_\Theta^{\prime\,2}+\Gamma_\Theta^2}\right)\cos\delta\right.\\
\n\\
\n
&&+\left.\left(R\frac{\Delta_\phi\Gamma_\Theta-\Gamma_\phi\Delta_\Theta}
{\Delta_\Theta^2+\Gamma_\Theta^2}
+R'\frac{\Delta_\phi\Gamma_\Theta-\Gamma_\phi\Delta'_\Theta}
{\Delta_\Theta^{\prime\,2}+\Gamma_\Theta^2}\right)\sin\delta\right].
\eea
The three pairs' invariant masses are constrained by
$m_{12}^2+m_{13}^2+m_{23}^2=s+2m_{K^0}^2+m_p^2$; hence the three small
deviations from the resonance centers are constrained by
\beq
\Delta_\Theta m_\Theta+\Delta'_\Theta m_\Theta+\Delta_\phi m_\phi
=\Delta_\gamma m_p
\la{deltas}\eeq
where $\Delta_\gamma$ is the (doubled) deviation of the photon lab energy
from $1.87\,{\rm GeV}$ where all three resonance strips cross at one point
in the Dalitz plot:
\bea\la{Delta-gamma}
&&\Delta_\gamma:=2(E_\gamma-E_\gamma^{(3)}),\\
\n\\
\n
&&E_\gamma^{(3)}:=\frac{2m_\Theta^2+m_\phi^2-2m_{K^0}^2-m_p^2}{2m_p}
\approx 1.87\,{\rm GeV}.
\eea
\begin{figure}[htb]
\includegraphics[width=0.22\textwidth]{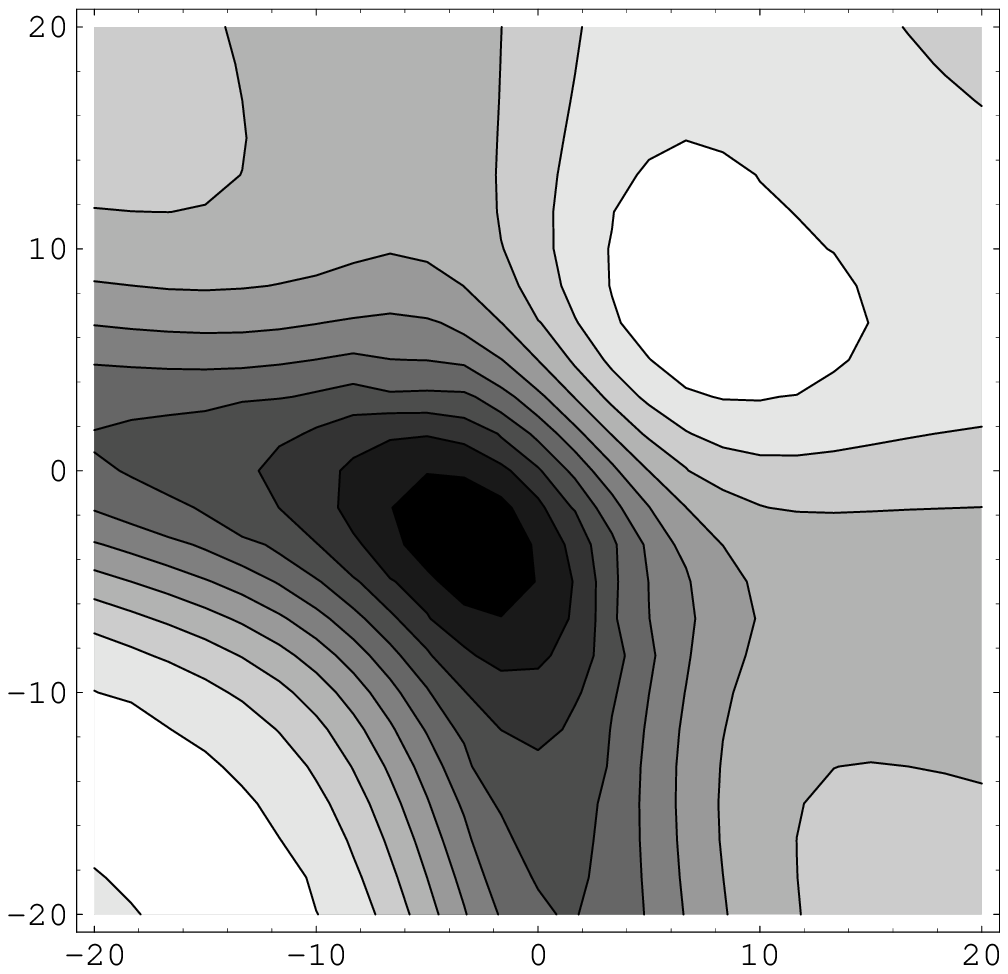}
\includegraphics[width=0.22\textwidth]{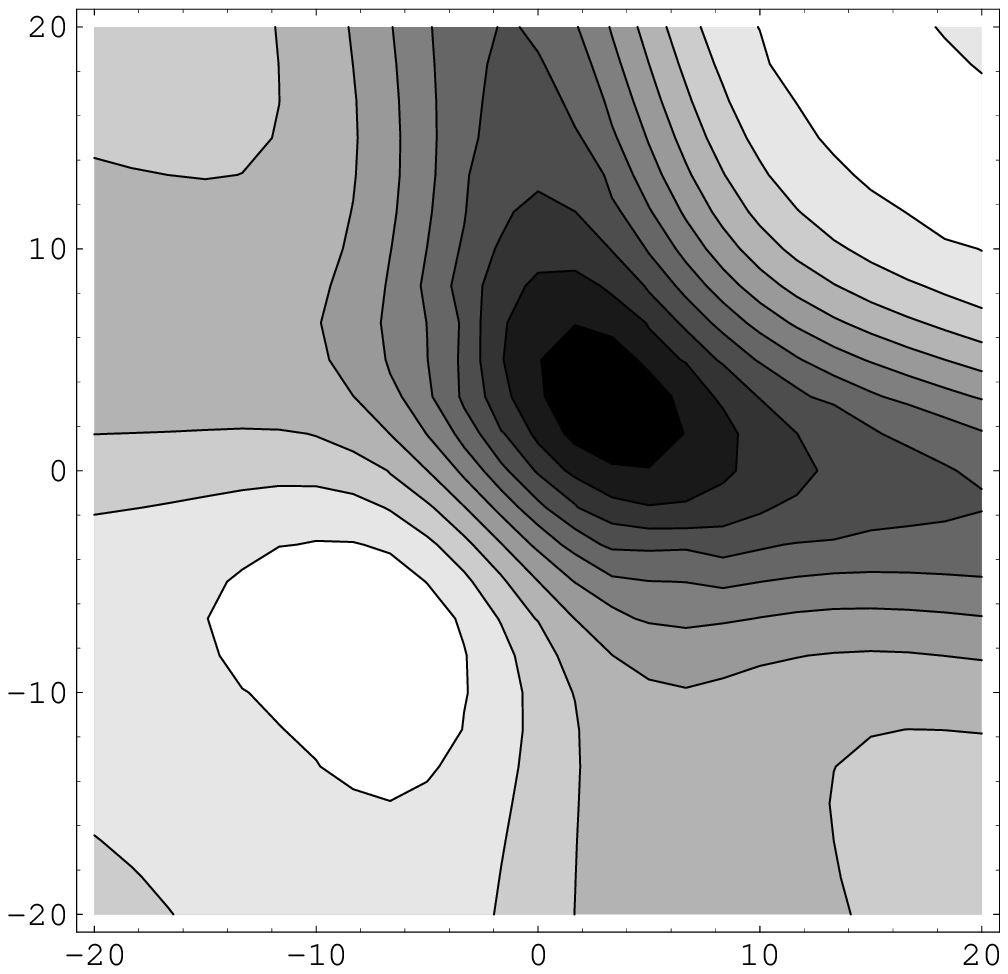}
\caption{Triple $\phi\!-\!\Theta(K_{\!S}p)\!-\!\Theta(K_{\!L}p)$ interference pattern
from \eq{sigma3}, plotted in the $m_{13}\!=\!m(K_{\!L}p)$ (horizontal) and $m_{23}\!=\!m(K_{\!S}p)$
(vertical) axes, in MeV. The axes' zeroes correspond to $m_{13}=m_{23}=m_\Theta$.
The left panel shows the excess (darker) and deficiency (lighter) of events
integrated over the range of $E_\gamma$ from 1.82 to 1.87 GeV, smeared with
the experimental resolution of 5 MeV. The right panel shows the events
integrated over the range of $E_\gamma$ from 1.87 to 1.92 GeV. The relative
phase $\delta$ of the $\phi$ and $\Theta^+$ production amplitudes is assumed to be zero.}
\end{figure}

Note that, although the $\Theta$ amplitude is antisymmetric under the interchange
of $K_{\!S}\leftrightarrow K_{\!L}$, so is the $\phi$ amplitude, therefore the
interference cross section is symmetric, hence the relative plus sign in the $R,R'$ terms.
In the ideal acceptance case the ratios of the production amplitudes are equal,
$R=R'$. In reality, however, these ratios may appear to be non-equal
owing to different ways one registers $K_{\!S},K_{\!L}$.

\Eq{sigma3} predicts a rich structure of event density in the Dalitz plot
near the triple interference point at $E_\gamma\approx 1.87\,{\rm GeV}$, which,
however, needs to be smeared by the experimental resolution. We assume that
errors in measuring the invariant masses $m_{13}$ and $m_{23}$ are
statistically independent and are given by the Gaussian distributions with
dispersion $\sigma$ which we take equal 5 MeV. The interference pattern
survives smearing: it is presented in Fig.~7 which demonstrates a striking
asymmetry between event patterns above the energy $E_\gamma= 1.87\,{\rm GeV}$,
and below it.

Another way to stress the interference phenomenon (and also to reduce considerably
the background) is to integrate events in the upper right corner of Fig.~7 and
{\em subtract} events in the lower left corner, as function of the incident photon
energy. The resulting excess/deficiency as function of $E_\gamma$ is shown in Fig.~8.

\begin{figure}[htb]
\includegraphics[width=0.35\textwidth]{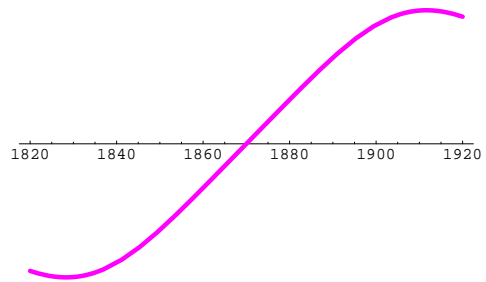}
\caption{Triple $\phi\!-\!\Theta(K_{\!S}p)\!-\!\Theta(K_{\!L}p)$ interference:
number of events with $m_{13},m_{23}<m_\Theta$ subtracted from the number
of events with $m_{13},m_{23}>m_\Theta$, as function of the photon lab
energy $E_\gamma$, in MeV.}
\end{figure}

Using \eq{sigma3} one can extract other peculiar characteristics of the
unique triple interference in the $\gamma p\to K_{\!L} K_{\!S}p$ reaction.

\section{$\Theta^+$ interference with $\Lambda(1520)$}

Interference can be probably observed also in the $\gamma p\to
K^+\Lambda(1520)\to K^+(\bar K^0n)$ reaction at photon lab energies
$E_\gamma$ between 1.81 and 2.17 GeV where the final state can interfere
with the same final state from the $\gamma p\to \bar K^0\Theta^+ \to
\bar K^0 (K^+n)$ reaction, see the phase volume plots in Fig.~9.

\begin{figure}[htb]
\includegraphics[width=0.22\textwidth]{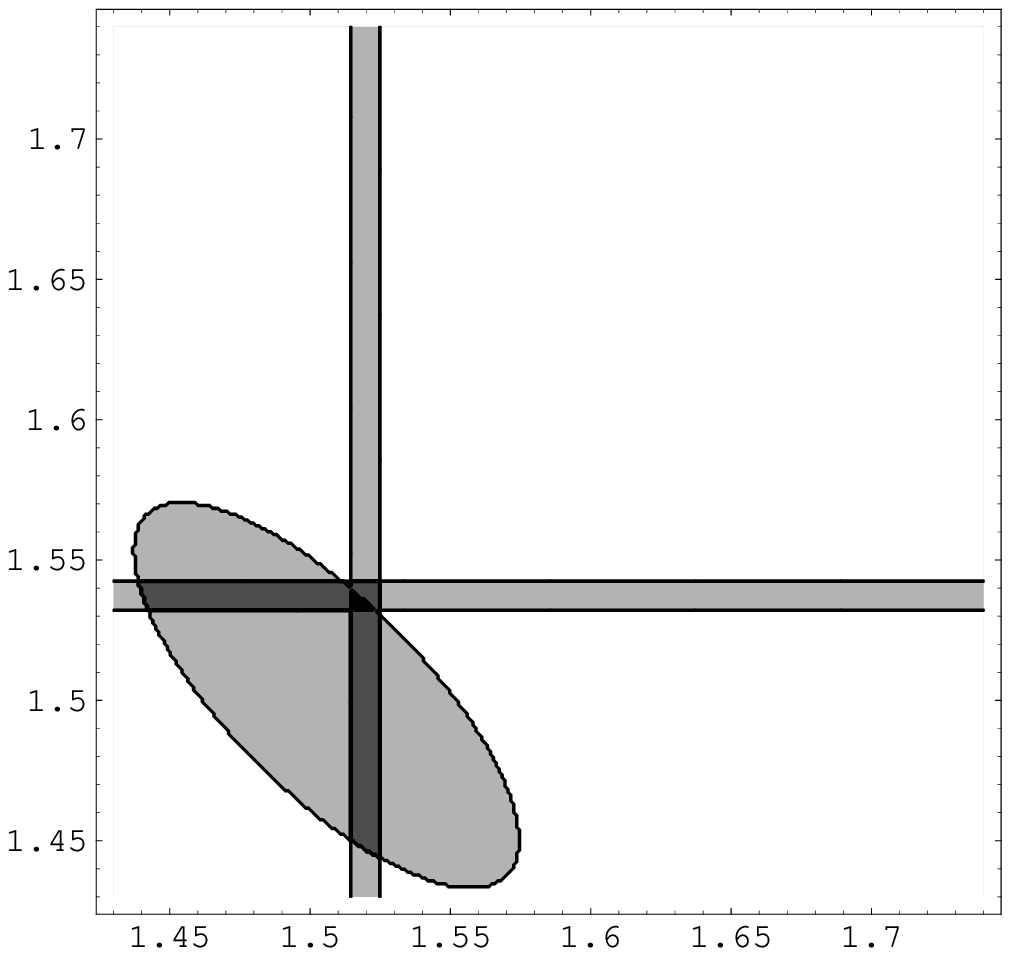} 
\includegraphics[width=0.22\textwidth]{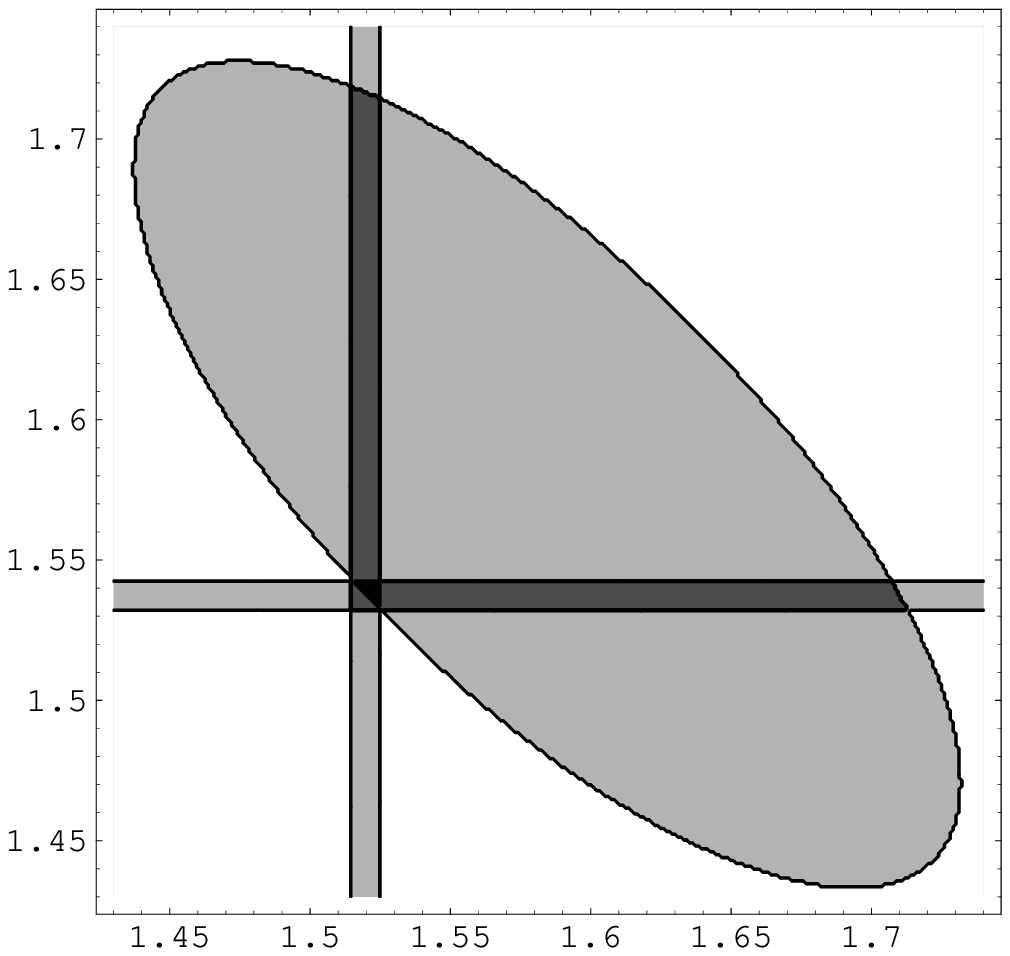} 
\includegraphics[width=0.22\textwidth]{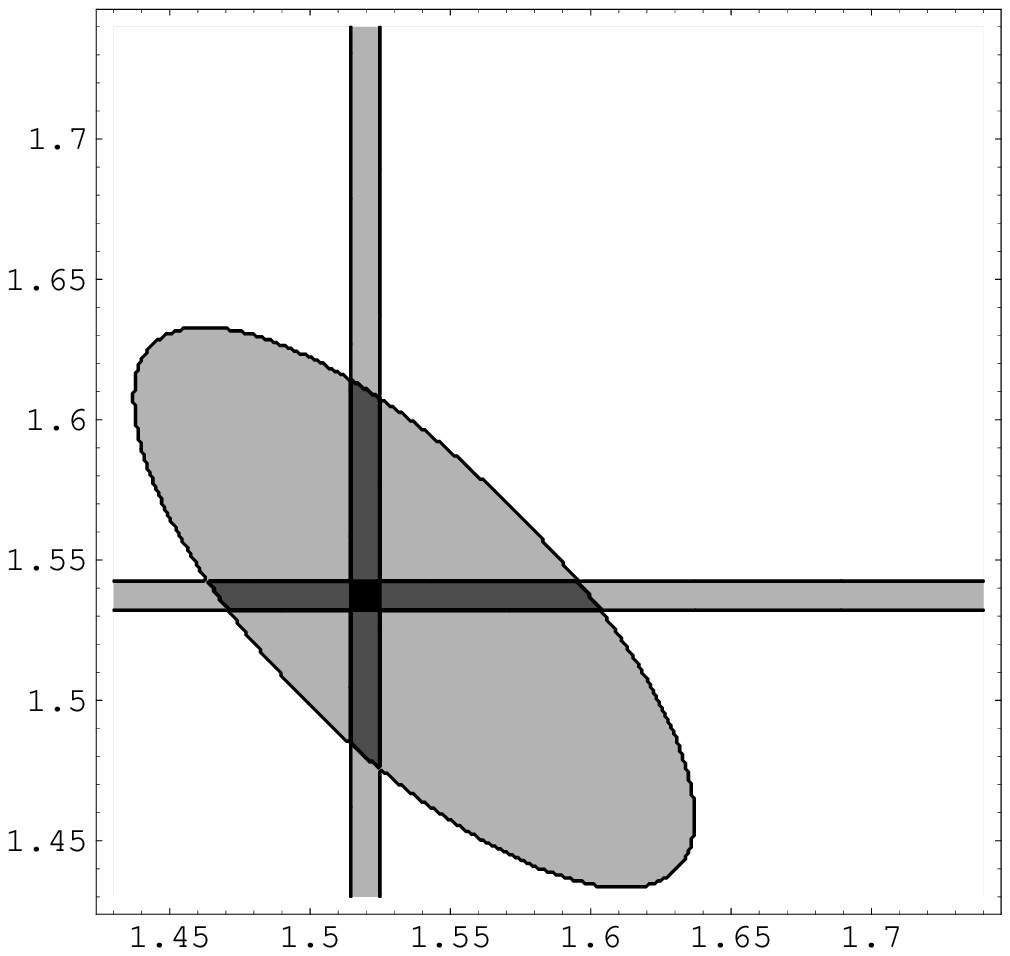} 
\caption{Phase volumes (Dalitz plots) for the $\gamma p\to \bar K^0K^+n$
reaction at photon lab energies $E_\gamma=1.81,\,2.17$ and $1.95\,{\rm GeV}$.
The $\bar K^0n$ invariant mass $m_{12}$ is plotted along the horizontal axis,
and the $K^+n$ invariant mass $m_{23}$ is plotted along the vertical axis
(note the linear scale). The strips show the positions of the $\Lambda(1520)$
and $\Theta^+$ resonances ($m_\Theta=1537\,{\rm MeV}$ is assumed).
The $\Theta\!-\!\Lambda$ interference occurs at the intersection of the strips,
thus within the range of $E_\gamma$ from $1.81$ to $2.17\,{\rm GeV}$.}
\end{figure}

This reaction has been studied in the same experiment~\cite{CLAS-p2}, also with
a null result for the $\Theta^+$ search. There were many final states found identified
with the $\bar K^0n$ decay of the $\Lambda(1520)$ resonance, however they were
cut out from the analysis~\cite{CLAS-p2}. We suggest that precisely these events
should be analyzed with respect to the possible interference with the $\Theta^+$ production.
The procedure to isolate the interference term is exactly as described above
for the case of the $\phi$ meson, the only difference being that
$\Lambda(1520)$ has the width larger than that of $\phi$, $\Gamma_\Lambda=15.6\,{\rm
MeV}$, therefore the interference picture may be less pronounced. However,
$\Gamma_\Lambda$ is still much less than the typical hadron masses determining
the scale of variation of the production amplitudes, therefore one may hope
that the equations of the previous Sections apply to this case as well.
In particular, subtracting events slightly below the $\Lambda$ resonance
from those slightly above, should result in the same \eq{sigmaassmea}
exhibiting an oscillating behavior in the $(K^+n)$ invariant mass
(but with a different background contribution denoted in \eq{sigmaassmea}
as $\rho\cos\eta$). The excess/deficiency of events due to the
$\Theta\!-\!\Lambda$ \inte is shown in Fig.~10 which is similar to Fig.~5.

\begin{figure}[htb]
\includegraphics[width=0.28\textwidth]{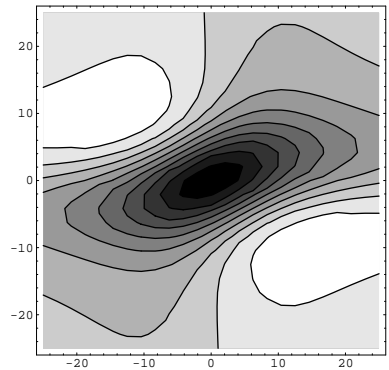}
\caption{Contour plot for the excess/deficiency of events smeared with the
experimental resolution of 5 MeV, which is due to the $\Lambda(1520)\!-\!\Theta$ interference,
\eq{inter2}. The excess is shown darker and the deficiency shown lighter.
$m(\bar K^0n)\!-\!m_\Lambda$ is plotted along the horizontal axis, and
$m(K^+n)\!-\!m_\Theta$ is plotted along the vertical axis, both in MeV. The relative
phase $\delta$ of the two production amplitudes is assumed to be zero.}
\end{figure}

\section{Conclusions}

Lately, strong signals of the exotic $\Theta^+$ baryon have been observed in
the direct formation experiment by the DIANA collaboration~\cite{Dolgolenko-2}
and in a quasi-formation experiment by the LEPS collaboration~\cite{Nakano-2}.
The results from numerous production experiments are still controversial,
the main reason being the small couplings of $\Theta^+$ with ordinary hadrons.
The origin of those small couplings are theoretically well understood
(see a recent brief review in Ref.~\cite{D-06}), however they preclude an easy
observation of the $\Theta^+$ in most of the production experiments.

To override the smallness of the $\Theta^+$ production cross sections, a
new way of $\Theta^+$ searching is suggested, based on the
interference between the small $\Theta^+$ production amplitude and the
large production amplitude of a known resonance yielding the same final state.
Owing to the narrowness of the resonances, a model-independent coherent
sum of the two resonance amplitudes can be applied giving an unambiguous
interference term. The interference may be substantial in the kinematical
range where the invariant masses of the resonances' decay products are close
to the resonances' centers. The pattern of events due to interference allows
to determine the relative phase of the two resonance amplitudes.

We have suggested a procedure for the analysis of the interference, based
on the {\em subtraction} of events above and below one of the resonances,
to purify the signature of the other one and to get rid of a large, if not
dominant, part of the background. Smearing by a finite experimental resolution
is not likely to blur out the characteristic oscillating nature of the
resulting event pattern. With sufficient statistics, the mass of the
$\Theta^+$ resonance can be accurately established and a tight upper limit
for its width determined.

The method is directly applicable to the $\gamma p \to \bar K^0K^0p$ and
$\bar K^0K^+n$ reactions at relatively low energies, studied in a recent CLAS
experiment. In the first case one should look for the $\Theta\!-\!\phi$ and in the
second case for the $\Theta\!-\!\Lambda(1520)$ interference. In the first reaction
also a unique {\em triple} interference can be examined at a particular
photon energy $E_\gamma\approx 1.87\,{\rm GeV}$.\\

The suggested method of studying $\Theta^+$ production through interference
is applicable to other experiments, wherever the $\Theta^+$ can kinematically
interfere with a known resonance whose production rate is large.

\newpage
{\bf Acknowledgements}
\vskip 0.2true cm

We are grateful to Ya.~Azimov, V.~Petrov and A.~Titov for useful discussions.
The work of M.A. is supported by the US DOE grant FG02-96ER40960.
D.D. acknowledges support by the A.v.Humboldt Research Award during the
visit to Bochum University. D.D.'s work is supported in part by Russian
Government grants 1124.2003.2 and RFBR 06-02-16786. M.P. is supported
by the Sofja Kowalewskaja Programme of the A.v.Humboldt Foundation,
by German Federal Ministry for Education and Research (BMBF), and DFG grant SFB TR-16.


\begin{thebibliography}{100}

\bibitem{Nakano-1}
T. Nakano [LEPS Collaboration], Talk at the PANIC 2002 (Oct. 3, 2002, Osaka);
T. Nakano {\it et al.}, Phys. Rev. Lett. {\bf 91}, 012002 (2003), arXive:
hep-ex/0301020.

\bibitem{Dolgolenko-1}
V.A. Shebanov [DIANA Collaboration], Talk at the Session of the Nuclear Physics
Division of the Russian Academy of Sciences (Dec. 3, 2002, Moscow); V.V. Barmin
{\it et al.}, Phys. Atom. Nucl. {\bf 66}, 1715 (2003) [Yad. Fiz. {\bf 66}, 1763
(2003)], arXive: hep-ex/0304040.

\bibitem{footnote-1}
The two experiments were totally independent as both groups didn't know about
the work of one another and made a tedious re-analysis of data taken long before,
however both searches were triggered off by the authors of Ref.~\cite{DPP-97}
where the resonance at $\sim 1530\,{\rm MeV}$ and width less than $15\,{\rm MeV}$
had been predicted.

\bibitem{DPP-97}
D.~Diakonov, V.~Petrov and M.~Polyakov, Z.~Phys. {\bf A359}, 305 (1997),
arXive: hep-ph/9703373; arXive: hep-ph/0404212.

\bibitem{CLAS-d2}
B. McKinnon {\it et al.} [CLAS Collaboration], Phys. Rev. Lett. {\bf 96},
212001 (2006), arXive: hep-ex/0603028.

\bibitem{CLAS-d2Lambda}
S.~Niccolai {\it et al.} [CLAS Collaboration], Phys. Rev. Lett. {\bf 97},
032001 (2006), arXive: hep-ex/0604047.

\bibitem{CLAS-p2}
R.~De Vita {\it et al.} [CLAS Collaboration], Phys. Rev. {\bf D74}, 032001
(2006), arXive: hep-ex/0606062.

\bibitem{Titov}
A.~Titov, B.~Kampfer, S.~Date and Y.~Ohashi, Phys. Rev. {\bf C72}, 035206
(2005), Erratum: ibid. {\bf C72}, 049901 (2005), arXive: nucl-th/0506072;
nucl-th/0607054.

\bibitem{Guzey}
V.~Guzey,
Phys. Rev. C {\bf 69}, 065203 (2004), arXive: hep-ph/0402060; arXive: hep-ph/0608129.

\bibitem{K*exchange}
H.~Kwee, M.~Guidal, M.~Polyakov and M.~Vanderhaeghen, Phys. Rev. {\bf D72},
054012 (2005), arXive: hep-ph/0507180.

\bibitem{D-06}
D.~Diakonov, talks at {\it Quarks 2006} (St.~Petersburg, May 19-26, 2006)
and at {\it Quark Confinement and Hadron Spectrum VII} (Ponta Delgada,
Sep. 2-7, 2006), to be published in the corresponding Proceedings, arXive:
hep-ph/0610166.

\bibitem{Nakano-2}
T.~Nakano, talk at the Bochum workshop on $\eta$ photoproduction (Feb. 23-25,
2006); talk at the Internat. Conf. on Strangeness in Quark Matter (UCLA, March
26-31, 2006) and other presentations.

\bibitem{Dolgolenko-2}
V.V.~Barmin {\it et al.} [DIANA Collaboration], arXive: hep-ex/0603017.

\bibitem{SVD-2}
A.~Kubarovsky, V.~Popov and V.~Volkov [for the SVD Collaboration], arXive:
hep-ex/0610050.

\bibitem{CT}
R.N.~Cahn and G.H.~Trilling, Phys. Rev. D {\bf 69}, 011501 (2004), arXive: hep-ph/0311245.

\bibitem{BELLE}
R.~Mizuk [for BELLE Collaboration], talk at the EPS International Europhysics
Conference on High Energy Physics (Lisbon, 21-27 Jul 2005), PoS HEP2005, 089
(2006).

\bibitem{Arndt:2003xz}
R.~A.~Arndt, I.~I.~Strakovsky and R.~L.~Workman,
Phys. Rev. C {\bf 68}, 042201 (2003)
[Erratum-ibid. {\bf 69}, 019901 (2004)], arXive: nucl-th/0308012;\\
A.~Sibirtsev, J.~Haidenbauer, S.~Krewald and U.~G.~Meissner,
Phys. Lett. B {\bf 599}, 230 (2004), arXive: hep-ph/0405099.

\bibitem{DP-05}
D.~Diakonov and V.~Petrov, Phys. Rev. {\bf D72}, 074009 (2005), arXive:
hep-ph/0505201.

\bibitem{Lorce}
C.~Lorc{\'e}, Phys. Rev. {\bf D74}, 054019 (2006), arXive: hep-ph/0603231.

\bibitem{Oganesian}
A.G.~Oganesian, arXive: hep-ph/0608031.

\bibitem{Polyakov-magnetic}
M.~Polyakov and A.~Rathke, Eur. Phys. J. {\bf A18}, 691 (2003),
arXive: hep-ph/0303138;\\
H.-C.~Kim {\it et al.}, Phys. Rev. {\bf D71}, 094023 (2005), arXive: hep-ph/0503237;\\
Ya.~Azimov, V.~Kuznetsov, M.~V.~Polyakov and I.~Strakovsky,
Eur. Phys. J. A {\bf 25} (2005) 325, arXive: hep-ph/0506236.

\bibitem{footnote-2}
This upper limit was obtained assuming certain extrapolation of the data
to the kinematical range experimentally unaccessable~\cite{CLAS-p2}. The actual
upper limit for the $\Theta^+$ production may be therefore somewhat larger.

\bibitem{Azimov:2006he}
Ya.~Azimov, V.~Kuznetsov, M.~Polyakov and I.~Strakovsky,
arXive: hep-ph/0611238.

\bibitem{Titov-05}
 T.~Mibe {\it et al.}  [LEPS Collaboration],
  Phys.\ Rev.\ Lett.\  {\bf 95} (2005) 182001
  [arXiv:nucl-ex/0506015].
\end{thebibliography}
\end{document}